\documentclass[%
 reprint,
 amsmath,amssymb,
 aps, prx,
floatfix,
]{revtex4-2}

\usepackage{dcolumn}

\usepackage[whole]{bxcjkjatype}

\usepackage{amsmath, amsfonts, amssymb, amsthm, ascmac, physics}
\usepackage{graphicx, xcolor}
\usepackage{comment}
\usepackage{appendix} 
\usepackage{hyperref}

\newtheorem*{theorem*}{Theorem}
\newtheorem*{lemma*}{lemma}

\begin{document}
\title{Reshaping Neural Representation via Associative, Presynaptic Short-Term Plasticity}
\author{Genki Shimizu}
\email[Corresponding author: ]{genki.shimizu@a.riken.jp}

\author{Taro Toyoizumi}
\email[Co-corresponding author: ]{taro.toyoizumi@riken.jp}
\affiliation{Laboratory for Neural Computation and Adaptation, RIKEN Center for Brain Science}
\affiliation{Graduate School of Information Science and Technology, University of Tokyo}

\date{\today}

\begin{abstract}
    
Short-term synaptic plasticity (STP) is traditionally viewed as a purely presynaptic mechanism that filters incoming spike trains without detecting pre-post correlations. Recent experiments, however, have revealed an associative form of STP in which presynaptic release probability is modified in concert with long-term potentiation, suggesting a richer computational role for presynaptic plasticity. Here, we develop a normative theory of such associative STP using an information-theoretic framework. Extending Fisher-information-based learning rules to synapses with Tsodyks–Markram dynamics, we derive analytic update rules for both postsynaptic weight and release probability that maximize encoded stimulus information. The resulting learning rules decompose into a conventional postsynaptic term that tracks local firing and a distinct presynaptic term. The latter term exhibits a novel phase-advanced structure relative to firing rate and selectively detects the onset of presynaptic activity; critically, it is within this presynaptic component that the plasticity of postsynaptic strength and release probability differ. When a neural population is sequentially activated on a timescale longer than the EPSP time constant, this onset sensitivity biases optimal connectivity toward anti-causal associations, strengthening synapses from neurons activated later in a stimulus to those activated earlier. In recurrent circuits, optimizing release probability together with postsynaptic strength yields larger gains in encoded stimulus information than weight-only learning and supports reverse replay sequences when external drive is removed. Furthermore, constraints on total release probability systematically modulate the strength of temporal asymmetry. Together, these results provide a principled account of the computational role of associative STP and highlight presynaptic plasticity of release probability as a key substrate for rapidly reconfigurable temporal coding.
\end{abstract}

\maketitle
\section{Introduction}

Learning through experience is mediated by synaptic plasticity, the activity-dependent modification of connection strengths between neurons. Classical long-term potentiation (LTP) has been extensively studied as the primary mechanism for associative learning, where correlated pre- and postsynaptic activity leads to persistent strengthening of synapses through increased postsynaptic receptor expression \cite{bliss1973Longlasting, bi1998Synaptic}. This Hebbian plasticity enables the formation of neural assemblies that encode relationships between concepts. However, LTP typically requires tens of minutes to hours for full expression and once expressed, can persist from hours to days or longer, which cannot fully account for flexible behavioral adaptation occurring on faster timescales.

Short-term plasticity (STP) operates on timescales from milliseconds to minutes, with short-term depression and facilitation modulating synaptic efficacy through presynaptic mechanisms—primarily vesicle depletion and calcium-dependent changes in release probability \cite{zucker2002Shortterm}. 
Importantly, classical STP has been considered a purely presynaptic phenomenon, governed solely by presynaptic firing history rather than by pre-post correlations.

Recently, Ucar et al. \cite{ucar2021Mechanical} discovered a novel form of associative short-term plasticity that challenges this traditional dichotomy. This plasticity requires coincident pre- and postsynaptic activity for induction—like LTP—but operates on rapid timescales characteristic of STP. The mechanism involves postsynaptic spine enlargement exerting mechanical pressure on presynaptic terminals, increasing vesicle release probability. This plasticity is induced within minutes and persists for tens of minutes to hours, potentially explaining online formation of associative memories at behavioral timescales \cite{ucar2021Mechanical, kasai2023Mechanical}.

What computational role might this “associative STP” play? A purely timescale-based argument—“it is just a fast form of LTP”—is unsatisfying, because LTP already has early phases and because the locus of expression matters: associative STP acts through presynaptic release probability and hence interacts with depletion dynamics in ways that classical, postsynaptic LTP does not. 
Hence, our question is: what computational features arise from associative learning rules that shape synaptic dynamics rather than static weights?

To address this question quantitatively, we adopt an information-theoretic framework. Neural populations have been shown to optimize their representations for efficient information transmission \cite{linsker1988Selforganization}, and Hebbian-like plasticity rules can emerge from that optimization process \cite{linsker1988Selforganization,toyoizumi2005Generalized, toyoizumi2005Spiketiming, toyoizumi2006Fisher, yoshida2023Information}.
While STP has been shown to enhance information processing in various contexts \cite{rotman2011Shortterm, regehr2012ShortTerm}, the impact of associative learning of STP parameters on neural encoding remains unexplored.
In this article, we focus on the functional significance of associative STP from the perspective of efficient neural representations. In particular, we use Fisher information as a local and tractable measure of encoding efficiency \cite{dayan2001Theoretical, brunel1998Mutual}.

To this end, we extend the Fisher information optimization framework \cite{seung1993Simple, toyoizumi2006Fisher} from static synaptic weights to dynamic synapses characterized by activity-dependent vesicle release. 
Specifically, we adopt the Tsodyks-Markram model \cite{tsodyks1997Neural, tsodyks1998Neural} of short-term synaptic dynamics and derive learning rules for both postsynaptic weight and release probability.

We find that short-term depression makes the presynaptic component of the learning rule onset-sensitive: presynaptic activity contributes most strongly when it begins, rather than during sustained firing. 
When the external stimulus drives a neural population slowly and sequentially, this onset sensitivity maximizes the overlap with postsynaptic activity for anti-causal pairings, where presynaptic neurons lag behind postsynaptic neurons.
As a result, optimizing synaptic efficacy preferentially strengthens \emph{anti-causal} connections, in sharp contrast to classical STDP.
In recurrent circuits, joint optimization of release probability and postsynaptic strength yields larger gains in encoded stimulus information than weight-only learning or static synapses, and the resulting backward connectivity supports reverse replay after the external drive is removed. 
Furthermore, the extent of temporal asymmetry depends on constraints on release probability, potentially explaining state-dependent differences in replay directionality observed during wakefulness versus sleep. 
Our framework provides a principled understanding of how associative short-term plasticity shapes neural representations through the interaction between correlation-based learning and short-term synaptic dynamics.

\section{Model}

\subsection{Neuron Model}

Following the setting of \cite{seung1993Simple, toyoizumi2006Fisher}, we consider a network of stochastically firing spiking neurons.

The membrane potential $u_i(t)$ of each neuron $i$ is given by the sum of stimulus-dependent external input $h_i(t, \theta)$ and recurrent synaptic inputs from other neurons:
\begin{equation}
u_i(t) = h_i(t, \theta) + \sum_{j=1}^N \sum_{t_j^{f_j}} \epsilon(t-t_j^{f_j}) w_{ij}(t_j^{f_j}),
\end{equation}
where $t_j^{f_j}$ denotes the time of the $f_j$-th spike of presynaptic neuron $j$, and $\sum_{t_j^{f_j}}$ denotes a sum over all of its spike times.
The time-dependent synaptic strength $w_{ij}(t)$ represents the amplitude of the excitatory postsynaptic potential (EPSP) generated at neuron $i$ by a spike from neuron $j$ arriving at time $t$. 
The kernel $\epsilon (t) = e^{-t/\tau_m}\Theta(t)$  describes a causal EPSP with membrane time constant $\tau_m$, where $\Theta(t)$ is the Heaviside step function.

For stochastic firing, we assume that the instantaneous firing rate $\rho_i(t)$ is determined by a nonlinear function $g(u)$ of the membrane potential, without considering refractory periods or membrane potential reset: $\rho_i(t) = g(u_i(t))$. Consequently, neuronal firing follows an inhomogeneous Poisson process. We primarily consider an exponential activation function $g(u) = g_c e^{\beta(u-u_c)}$, where $\beta$ controls the gain, and $u_c$ is the threshold. Our results remain qualitatively similar for other reasonable activation functions, such as the sigmoid function. This model can be viewed as a special case of the spike response model \cite{gerstner2014Neuronal}. Furthermore, we assume that spike generation is conditionally independent across neurons given the membrane potential and spike history, thereby neglecting common noise sources.

\subsection{Synapses with Short-Term Dynamics}
\label{subsec:STP-STDonly}
Synaptic strengths $w_{ij}(t)$ follow Tsodyks-Markram short-term plasticity dynamics \cite{tsodyks1997Neural, tsodyks1998Neural}. In this model, the synaptic efficacy is expressed as the product of a facilitation factor $u_{ij}(t)$ and a depression factor $d_{ij}(t)$:
\begin{equation}
w_{ij}(t) = w_{ij}^0 u_{ij}(t) d_{ij}(t)
\end{equation}
where $w_{ij}^0$ denotes the postsynaptic strength, $u_{ij}(t)$ corresponds to the vesicle release probability, and $d_{ij}(t)$ represents the fraction of available neurotransmitter resources. The variables $u_{ij}, d_{ij}$ are normalized such that $u_{ij}, d_{ij} \in [0, 1]$.

For simplicity, we treat the facilitation variable as constant in time: $u_{ij}(t) \equiv U_{ij}$.
This assumption is valid, e.g., when the facilitation time constant $\tau_f$ is sufficiently short compared to the typical inter-spike interval (see also the Discussion section).

Under this assumption, the dynamics of synaptic strength are governed solely by the evolution of the depression variable:
\begin{equation}
\dot{d}_{ij} = \frac{1-d_{ij}}{\tau_d} - U_{ij} d_{ij}(t^-) \delta(t - t^{\text{spike}})
\end{equation}
where $d_{ij}(t^-)$ denotes the value immediately before time $t$ (unaffected by spike input at time $t$), and $\delta(t)$ is the Dirac delta function representing presynaptic spikes.
The effective synaptic strength $w_{ij}(t) = w_{ij}^0 U_{ij} d_{ij}(t)$ then follows:
\begin{equation}
\dot{w}_{ij} = \frac{w_{ij}^0 U_{ij} - w_{ij}(t)}{\tau_d} - U_{ij} w_{ij}(t^-) \delta(t - t^{\text{spike}})
\end{equation}

In this formulation, the baseline strength $w_{ij}^0$ characterizes the postsynaptic component (e.g., receptor expression level), while the release probability $U_{ij}$ governs the presynaptic dynamics of neurotransmitter release.

\subsection{Fisher Information}
We consider a population of neurons receiving parameter-dependent input $h_i(t, \theta)$, where $\theta$ is the stimulus parameter to be encoded. 
$x_i(t) = \sum_{t_i^{f_i}} \delta(t - t_i^{f_i})$ denotes the output spike train of neuron $i$ 
and the complete spike history is denoted by $X(t) = \{x_i(t') \mid i = 1, \ldots, N; \, 0 \leq t' \leq t\}$.

As calculated in \cite{toyoizumi2006Fisher}, the Fisher information, which quantifies how accurately a downstream decoder can estimate $\theta$ from the population activity, is given by:
\begin{align}
    \label{eq:FI-def}
    J &= - \left \langle \frac{\partial ^2 \log P(X)(T)}{\partial \theta^2} \right \rangle _{X(T)} \\
      &=\int_0^T dt \sum_{i=1}^N \left\langle \left[ h'_i(t) \frac{g'_i(t)}{g_i(t)}\right]^2 \rho_i(t)\right\rangle_{X(t)}
\end{align}
where $h'_i(t) = \partial h_i(t, \theta)/\partial \theta$ represents the sensitivity of the external input to the parameter, $g'_i(t) = dg(u)/du|_{u=u_i(t)}$ is the derivative of the activation function evaluated at the current membrane potential, and $\langle \cdot \rangle_{X(t)}$ denotes the average over the stochastic spike history $X(t)$.

While \cite{toyoizumi2006Fisher} derived learning rules for time-invariant synaptic weights that maximize Fisher information, our goal is to extend this framework to synapses with short-term dynamics.
Specifically, we aim to derive learning rules for both the postsynaptic strength $w_{ij}^0$ and the release probability $U_{ij}$ that maximize the Fisher information in networks with dynamic synapses. Although the initial postsynaptic response scales with the overall product of $w_{ij}^0 U_{ij}$, these parameters play distinct roles: $U_{ij}$ additionally governs how synaptic efficacy evolves during presynaptic spike trains through STP dynamics, justifying their separate optimization.

\section{Results}
We now analyze how associative short-term synaptic plasticity shapes Fisher-information-maximizing learning rules. First, we derive an exact score-function/eligibility-trace expression for the gradient that is valid for arbitrary recurrent coupling and can be estimated from network simulations. We then obtain a weak-coupling reduction that yields a compact analytic learning rule, clarifying how short-term depression creates an onset-sensitive presynaptic component and biases learning toward anti-causal associations. Finally, we study a one-dimensional ring driven by traveling-wave input as a concrete case study and validate the weak-coupling predictions using the exact gradient estimator in fully recurrent networks.

\subsection{Exact gradient and simulation-based estimation}
\label{sec:derive-learning-rule}
We begin with an exact expression for the Fisher-information gradient with respect to synaptic parameters $Z_{ij}\in\{w_{ij}^0,U_{ij}\}$, and show how it can be estimated from network simulations via eligibility traces.

\subsubsection{Exact gradient for arbitrary recurrent coupling}
\label{subsec:grad-computation}

We first state an exact expression for the gradient of the Fisher information that
holds for arbitrary recurrent coupling strengths. 
Recall that, for a realized spike history $X(T)$, the pathwise contribution
whose expectation gives the Fisher information can be written with
\begin{align}
\label{eq:FI-functional}
\mathcal{J}[X]
:= \int_0^T dt \sum_{k=1}^N
\left[
h_k'(t,\theta)\frac{g_k'(t)}{g_k(t)}
\right]^2 \rho_k(t),
\end{align}
where $\rho_k(t)=g(u_k(t))$ and $g_k'(t)=\left.\frac{dg}{du}\right|_{u=u_k(t)}$
(and similarly for higher derivatives). Then, the Fisher information is $J=\langle \mathcal{J}[X]\rangle_{X(T)}$.

For any synaptic parameter $Z_{ij}\in\{w_{ij}^0,\,U_{ij}\}$, differentiating the expectation
$J(Z)=\langle \mathcal{J}[X]\rangle_{X(T)}$ yields the score-function identity
\begin{align}
\label{eq:general-grad-score-identity}
\frac{\partial J}{\partial Z_{ij}}
=
\left\langle
\frac{\partial \mathcal{J}[X]}{\partial Z_{ij}}
+
\mathcal{J}[X]\,
\frac{\partial}{\partial Z_{ij}}\log P(X(T)\mid Z)
\right\rangle_{X(T)} .
\end{align}
Because spikes are conditionally independent across neurons given the membrane
potentials and the spike history, and because $Z_{ij}$ affects the membrane
potential only through the postsynaptic neuron $i$, both terms in
\eqref{eq:general-grad-score-identity} can be expressed using an \emph{eligibility trace}
\begin{align}
\label{eq:eligibility-def}
e_{ij}^{Z}(t) := \frac{\partial u_i(t)}{\partial Z_{ij}} .
\end{align}
A direct differentiation of \eqref{eq:FI-functional} gives
\begin{align}
\label{eq:pathwise-term}
\frac{\partial \mathcal{J}[X]}{\partial Z_{ij}}
&=
\int_0^T dt\;
\rho_i(t)\,\eta_i(t)\,e_{ij}^{Z}(t),
\end{align}
where we introduced the information-weighting factor
\begin{align}
\label{eq:eta-general}
\eta_i(t)
:=
\left[
h_i'(t,\theta)\frac{g_i'(t)}{g_i(t)}
\right]^2
\left(
\frac{2g_i''(t)}{g_i'(t)}-\frac{g_i'(t)}{g_i(t)}
\right).
\end{align}
For the exponential nonlinearity $g(u)=g_c e^{\beta(u-u_c)}$, the ratio
$g'/g=\beta$ is constant and \eqref{eq:eta-general} simplifies to
\begin{align}
\label{eq:eta-exp}
\eta_i(t)=\beta^3\,h_i'(t,\theta)^2 .
\end{align}
Moreover, the score term admits the standard point-process form
\begin{align}
\label{eq:score-term}
\frac{\partial}{\partial Z_{ij}}\log P(X(T)\mid Z)
=
\int_0^T
dt \left[x_i(t)-\rho_i(t)\right]\,
\frac{g_i'(t)}{g_i(t)}\,
e_{ij}^{Z}(t).
\end{align}

Equations \eqref{eq:general-grad-score-identity}--\eqref{eq:score-term} provide an exact
representation of $\partial J/\partial Z_{ij}$ for finite recurrent coupling.
A complete derivation is given in Appendix~\ref{appx:general-gradient}.

\subsubsection{Computability from network simulations}
\label{subsec:grad-computability}

Importantly, the right-hand side of \eqref{eq:general-grad-score-identity} depends only on quantities available along a single simulated trajectory $X(T)$:
$\rho_i(t)$, $g_i'(t)/g_i(t)$, and the eligibility trace $e_{ij}^Z(t)$.
The trace $e_{ij}^Z(t)$ can be updated online during simulation by differentiating the
membrane-potential dynamics with respect to $Z_{ij}$.
Specifically, for the exponential kernel $\epsilon(t)=e^{-t/\tau_m}\Theta(t)$, $e_{ij}^Z(t)$ obeys a simple event-driven recursion:
it decays continuously between presynaptic spikes and jumps at $t=t_j^f$ proportional to $\partial_{Z_{ij}} w_{ij}(t_j^f)$, which is computed from the STP variables
(see Appendix~\ref{appx:eligibility-traces} for the explicit update formulae).
Thus, an unbiased Monte Carlo estimator of $\partial J/\partial Z_{ij}$ is obtained by
averaging \eqref{eq:general-grad-score-identity} over repeated simulations (or, in stationary/periodic
regimes, by time-averaging after transients).

\subsection{Weak-coupling reduction and interpretation}
\label{sec:weak-coupling}
To obtain analytic intuition for the structure of the learning rules, we next consider the weak-coupling regime
($w_{ij}(t)\ll 1$), where the Fisher-information gradient reduces to a tractable correlation between a postsynaptic
information-weighting factor and a filtered, STP-modulated presynaptic drive.

\subsubsection{Weak-coupling reduction}
\label{subsec:grad-weak-coupling}

While \eqref{eq:general-grad-score-identity} is exact, it typically requires Monte Carlo averaging.
Following \cite{toyoizumi2006Fisher}, we obtain a compact analytic form by assuming weak synaptic coupling,
$w_{ij}^0\ll 1$ and consequently $w_{ij}(t)\ll 1$, and expanding around the baseline state $w_{ij}^0\equiv 0$.
At $w=0$, the conditional intensity reduces to a deterministic function
\begin{align}
\nu_i^0(t):=g(h_i(t,\theta)),
\end{align}
and $\mathcal{J}[X]$ becomes non-random; consequently, the score term in \eqref{eq:general-grad-score-identity}
vanishes at leading order.
Using the compensation formula for inhomogeneous Poisson spiking at $w=0$
(Appendix~\ref{appx:weak-coupling-compensation}), the gradient reduces to
\begin{multline}
\label{eq:FI-grad}
\frac{\partial J}{\partial Z_{ij}}
= \int_0^T dt\, \nu_i^0(t)\eta_i(t) 
\int_0^t dt'\, \epsilon(t-t')\, \\
\frac{\partial \langle w_{ij}(t') \rangle_{X(t')}}{\partial Z_{ij}}\,
\nu_j^0(t')
\qquad (Z_{ij}\in\{w_{ij}^0,\,U_{ij}\}),
\end{multline}
where $\langle\cdot\rangle_{X(t)}$ denotes averaging over the baseline ($w=0$) spike statistics.

To characterize how short-term depression shapes the presynaptic contribution, we introduce normalized sensitivity functions
\begin{subequations}
\label{eq:def-fw0-fU}
\begin{align}
f^{w_0}_{ij}(t) &:= \frac{1}{U_{ij}} \frac{\partial \langle w_{ij}(t) \rangle_{X(t)}}{\partial w_{ij}^0}, \\
f^{U}_{ij}(t) &:= \frac{1}{w^0_{ij}} \frac{\partial \langle w_{ij}(t) \rangle_{X(t)}}{\partial U_{ij}} .
\end{align}
\end{subequations}
Under weak coupling, these functions obey closed dynamics induced by the STD model
(Appendix~\ref{appx-subsec:fw0fU-derivation}):
\begin{subequations}
\label{eq:fw0fU-dynamics}
\begin{align}
\dot{f}^{w_0}_{ij} &= \frac{1}{\tau_d} - \left( \frac{1}{\tau_d} + \nu_j^0(t) U_{ij} \right)f^{w_0}_{ij}(t), \\
\dot{f}^{U}_{ij} &= \frac{1}{\tau_d} - \left( \frac{1}{\tau_d} + \nu_j^0(t) U_{ij} \right)f^U_{ij}(t) - \nu_j^0(t) U_{ij} f^{w_0}_{ij}(t).
\end{align}
\end{subequations}
Equation~\eqref{eq:fw0fU-dynamics} exposes a qualitative asymmetry between learning $w_0$ and learning $U$. 
A perturbation of $w_0$ rescales the effective synaptic efficacy without changing the depression dynamics themselves. 
By contrast, a perturbation of $U$ has two opposing effects: 
it directly increases release at the current presynaptic spike, 
but also accelerates depletion of synaptic resources.
The last term in the $f^U$ equation, $-\nu_j^0(t)U_{ij}f^{w_0}_{ij}(t)$, 
is the differential signature of this self-limiting effect.
Thus increasing $U$ at times when presynaptic activity is not informative can deplete resources and reduce transmission at later, more informative times.

Substituting \eqref{eq:def-fw0-fU} into \eqref{eq:FI-grad} yields
\begin{subequations}
\label{eq:FI-grad-using-f}
\begin{align}
\frac{\partial J}{\partial w_{ij}^0}
&=
U_{ij}\int_0^T dt\, \nu_i^0(t)\eta_i(t)
\int_0^t dt'\, \epsilon(t-t') f^{w_0}_{ij}(t') \nu_j^0(t'), \\
\frac{\partial J}{\partial U_{ij}}
&=
w_{ij}^0\int_0^T dt\, \nu_i^0(t)\eta_i(t)
\int_0^t dt'\, \epsilon(t-t') f^{U}_{ij}(t') \nu_j^0(t').
\end{align}
\end{subequations}

\subsubsection{``Hebbian'' factorization and implications for learning}
\label{subsec:hebbian-form}

The weak-coupling gradient \eqref{eq:FI-grad-using-f} admits a compact ``Hebbian'' form as a product of pre- and postsynaptic components:
\begin{align}
\label{eq:FI-grad-Hebbform}
    \frac{\partial J}{\partial Z_{ij}}
    &\propto \int_0^T dt \,
    \underbrace{C_{\mathrm{post},i}(t)}_{\text{postsynaptic}}
    \underbrace{\left[ \epsilon(t) * C_{\mathrm{pre},ij}^{Z}(t) \right]}_{\text{presynaptic}},
\end{align}
where $Z$ represents either $w_0$ or $U$. Here
\begin{align}
\label{eq:Cpost-def}
    C_{\mathrm{post},i}(t) := \nu_i^0(t)\eta_i(t),
\end{align}
and 
\begin{align}
\label{eq:Cpre-def}
    C_{\mathrm{pre},ij}^{Z}(t) := f^{Z}_{ij}(t)\nu_j^0(t).
\end{align}
Thus, the learning rule detects the correlation between the postsynaptic information-bearing component
$C_{\mathrm{post},i}(t)$ and a filtered version of the effective presynaptic drive
$C_{\mathrm{pre},ij}^{Z}(t)$.

\begin{figure}[!htbp]
    \centering
    \includegraphics[width=0.9\linewidth]{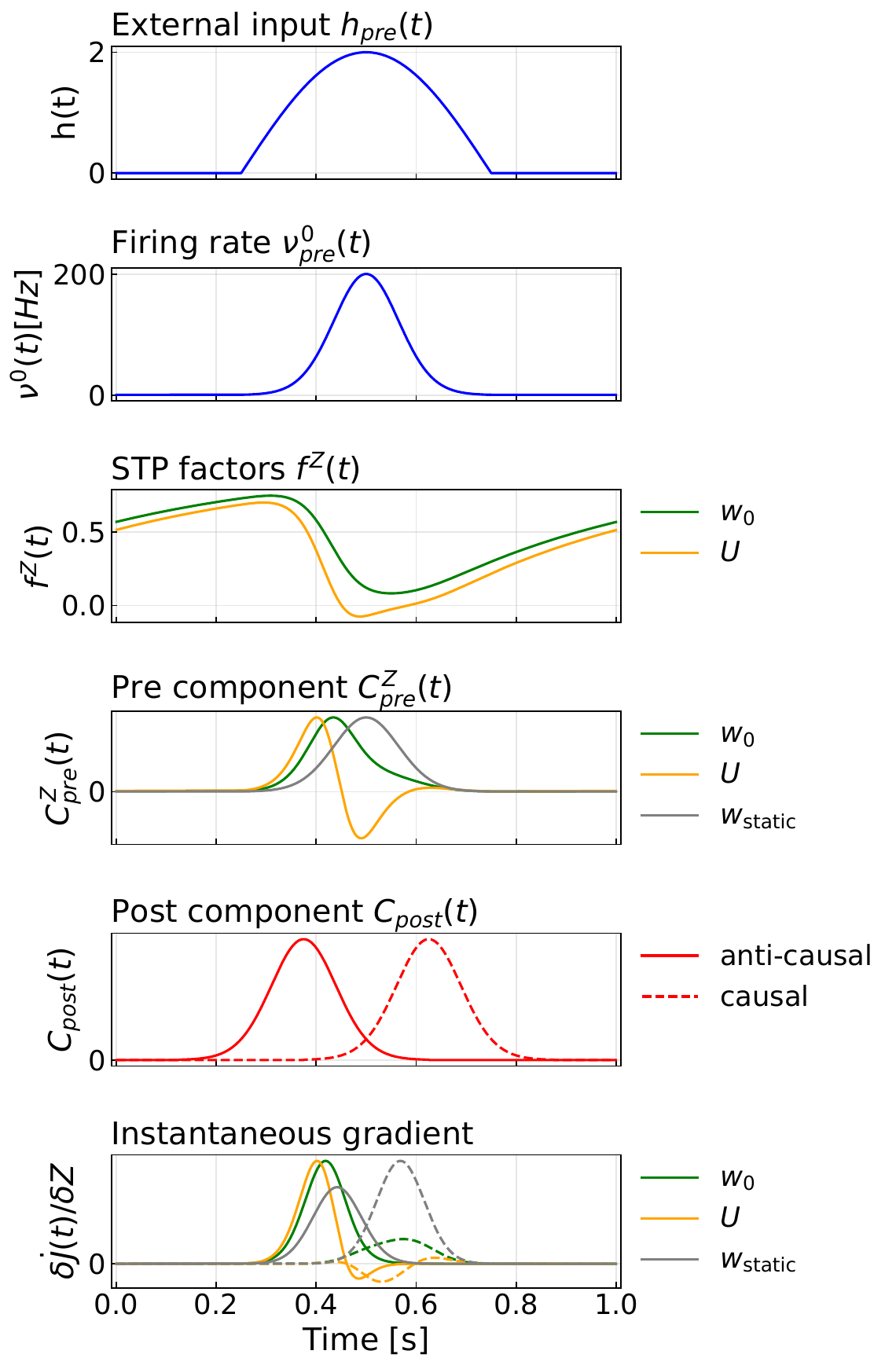}
    \caption[Pre- and postsynaptic components in the weak-coupling learning rule.]{
    \textbf{Pre- and postsynaptic components in the weak-coupling learning rule.}
    Example decomposition of the gradient into the postsynaptic component
    $C_{\mathrm{post}}(t)=\nu^0_i(t)\eta_i(t)$ and the presynaptic components
    $C_{\mathrm{pre}}^{Z}(t)=f^{Z}(t)\nu^0_j(t)$ ($Z\in\{w_0,U\}$) for a representative pair of neurons activated
    sequentially by a common stimulus (here, the traveling-wave input used in Sec.~\ref{subsec:STP-traveling-wave}).
    Short-term depression makes $C_{\mathrm{pre}}^{Z}$ transient and onset-weighted; for $Z=U$, $C_{\mathrm{pre}}^{U}$
    can become negative during sustained activation because increasing release probability accelerates depletion.
    The contribution to the Fisher-information gradient is controlled by the temporal overlap between
    $C_{\mathrm{pre}}^{Z}$ and $C_{\mathrm{post}}$; the overlap is larger for the \emph{anti-causal} pairing than for the
    \emph{causal} pairing.
    Colors indicate $Z$ (green: $w_0$, yellow: $U$); gray denotes the corresponding static-synapse case in the
    \emph{Pre component} and \emph{Instantaneous gradient} panels.
    In the \emph{Post component} and \emph{Instantaneous gradient} panels, solid lines correspond to
    \emph{anti-causal} activation (postsynaptic precedes presynaptic), whereas dotted lines correspond to
    \emph{causal} activation (presynaptic precedes postsynaptic).
    Parameters: $\beta = 2.0$, $g_c = 10.0$, $u_c = 1.0$, $\tau_d = 0.5~\mathrm{s}$,  $\tau_s = 0.01~\mathrm{s}$,
     $A = 3.0$, $\theta_c = \pi/2$, $\omega = 2\pi$.
    }
    \label{fig:pre-post_factors}
\end{figure}

Because STD induces depression during sustained presynaptic firing, the sensitivities $f^Z_{ij}(t)$ decrease with recent presynaptic activity, hence $C_{\mathrm{pre},ij}^{Z}(t)=f^Z_{ij}(t)\nu_j^0(t)$ becomes transient and emphasizes increases (onsets) in presynaptic activity (Figure~\ref{fig:pre-post_factors}). 
Particularly for $Z=U$, the presynaptic component $C_{\mathrm{pre},ij}^{U}(t)$ shows more pronounced onset sensitivity and can even become negative during sustained activation (Figure~\ref{fig:pre-post_factors}), because increasing release probability accelerates depletion and can suppress transmission at later times.
In Appendix~\ref{appx-sec:std-sensitivity-response}, we provide a more detailed analysis of the dynamics of $C_{\mathrm{pre},ij}^{U}(t)$ in response to both a step increase and weak sinusoidal modulation of $\nu_j^0(t)$ as tractable examples.

Consequently, the learning rule reinforces connections where the onset of presynaptic activity (large
$C_{\mathrm{pre}}^{Z}$) predicts high postsynaptic information (large $C_{\mathrm{post}}$). 
In sequentially activated populations, this yields \emph{anti-causal} associations, i.e., connections from neurons that fire later to neurons
that fired earlier in the stimulus-evoked sequence (Figure~\ref{fig:pre-post_factors}). 
This tendency occurs for both $U_{ij}$ and $w^0_{ij}$, 
but the stronger onset/phase lead of $C_{\mathrm{pre}}^{U}$ suggests that optimizing release probability should make the anti-causal bias particularly pronounced.

\subsection{Case Study: Traveling Waves on a Ring}
\label{subsec:STP-traveling-wave}
To better understand the circuit-level consequences of the learning rules, we consider neurons arranged on a one-dimensional ring receiving traveling-wave input.
In this case study, we first use the weak-coupling approximation to obtain analytic intuition for the phase structure of the Fisher-information gradients under rotational symmetry.
We then quantify learning-induced changes in neural representations and Fisher information in the full recurrent network using the exact (simulation-based) gradient estimator derived in Sec.~\ref{sec:derive-learning-rule}.

\subsubsection{Settings}
We consider neurons uniformly distributed on a ring, taking the limit $N \rightarrow \infty$. Each neuron is specified by its angular position $z$ on the ring, which physiologically corresponds to its preferred orientation.
Neurons receive traveling-wave input $h(z, t) = h(\omega t - z)$ with constant velocity and shape. 
We assume that the learned parameters possess rotational symmetry, such that $w_0(z, z') = w_0(\Delta z)$ and $U(z, z') = U(\Delta z)$, where $\Delta z = z - z'$ represents the phase difference between presynaptic position $z'$ and postsynaptic position $z$.

In the following analysis, we specifically consider the input $h(z, t) = A[\cos(\omega t - z) - \cos\theta_c]_+$ as a simple model of a localized, feature-selective drive.
This input has:
\begin{itemize}
    \item Spatially: a bump of width $2\theta_c$ centered at $z = \omega t$, rotating with angular velocity $\omega$
    \item Temporally: a bump of duration $2\theta_c/\omega$ centered at $t = (z + 2\pi n)/\omega$, with frequency $\omega/2\pi$
\end{itemize}
The encoded parameter is $\theta_c$, which controls the locality and selectivity of the external input.

For finite recurrent coupling, the microscopic gradients $\partial J/\partial Z(z,z')$
(with $Z\in\{w_0,U\}$) can be estimated from network simulations via the exact score-function/eligibility-trace identity
(Eq.~\eqref{eq:general-grad-score-identity}).
Rotational symmetry then implies that the functional derivative with respect to the profile $Z(\Delta z)$
is obtained by averaging over all pre--post pairs that share the same phase offset:
\begin{equation}
\label{eq:FI-grad-circular-exact}
\frac{\delta J}{\delta Z(\Delta z)}
=
\frac{1}{2\pi}\int_{-\pi}^{\pi} dz\;
\left.\frac{\partial J}{\partial Z(z,z')}\right|_{z'=z-\Delta z}
\quad (Z=w_0,U).
\end{equation}
We use this exact estimator when assessing the learning-induced changes in network representations and information transmission.

To obtain a transparent expression for the phase dependence of the gradients,
we also consider the weak-coupling approximation developed above.
In this limit, the profile gradients reduce to:
\begin{subequations}
    \label{eq:FI-grad-circular}
    \begin{multline}
            \frac{\delta J}{\delta w_0(\Delta z)}
            = U(\Delta z) \int_{-\pi}^\pi \frac{1}{2\pi} dz \int_0^T dt\;
            \nu^0(z, t)\eta^0(z, t) \\
            {}\times
            \int_0^t dt' \;
            \epsilon(t-t')\, f^{w_0}(t'; z, z-\Delta z)\, \nu^0(z-\Delta z, t'),
    \end{multline}
    \begin{multline}        
            \frac{\delta J}{\delta U(\Delta z)}
            = w_0(\Delta z) \int_{-\pi}^\pi \frac{1}{2\pi} dz \int_0^T dt\;
            \nu^0(z, t)\eta^0(z, t) \\
            {} \times
            \int_0^t dt' \;
            \epsilon(t-t')\, f^{U}(t'; z, z-\Delta z)\, \nu^0(z-\Delta z, t').
    \end{multline}
\end{subequations}
Here $\nu^0(z,t)=g(h(z,t))$, and $\eta^0(z,t)$ is the corresponding information-weighting factor
(Eq.~\eqref{eq:eta-general} evaluated at $u=h$).
The auxiliary functions $f^{w_0}(t;z,z')$ and $f^{U}(t;z,z')$ are the weak-coupling sensitivity functions
(Eq.~\eqref{eq:def-fw0-fU}) and obey the same dynamics as Eq.~\eqref{eq:fw0fU-dynamics} with the substitutions
$\nu_j^0(t)\to \nu^0(z',t)$ and $U_{ij}\to U(z-z')$.

\subsubsection{Weak-coupling prediction: asymmetric optimal profiles}
\label{subsec:weak-coupling-prediction}
In this subsection, we use the \emph{weak-coupling} gradients in Eq.~\eqref{eq:FI-grad-circular} to predict how the
optimal profiles $U(\Delta z)$ and $w_0(\Delta z)$ depend on the phase offset $\Delta z$ under resource constraints.

Under the same weak-coupling approximation, we can display the optimal profiles explicitly by close analogy with \cite{toyoizumi2006Fisher}.
For $w_0(\Delta z)$, we impose a zero-mean (balance) constraint and a fixed variance (synaptic cost), which yields an optimal profile that is proportional to the gradient shape:
\begin{align}
    w_0(\Delta z) \propto \frac{\delta J}{\delta w_0(\Delta z)},
\end{align}
up to an additive constant and overall scaling set by the constraints.
For $U(\Delta z)$, we impose a fixed mean release-probability budget
$\langle U(\Delta z) \rangle=\bar U$
and the bound $U(\Delta z)\in[0,1]$, so that the optimum is selected by the level sets of
$\delta J/\delta U(\Delta z)$ under this budget (Fig.~\ref{fig:grad-U}, white contours).

\begin{figure*}[!htbp]
    \centering
    \includegraphics[width=0.8\linewidth]{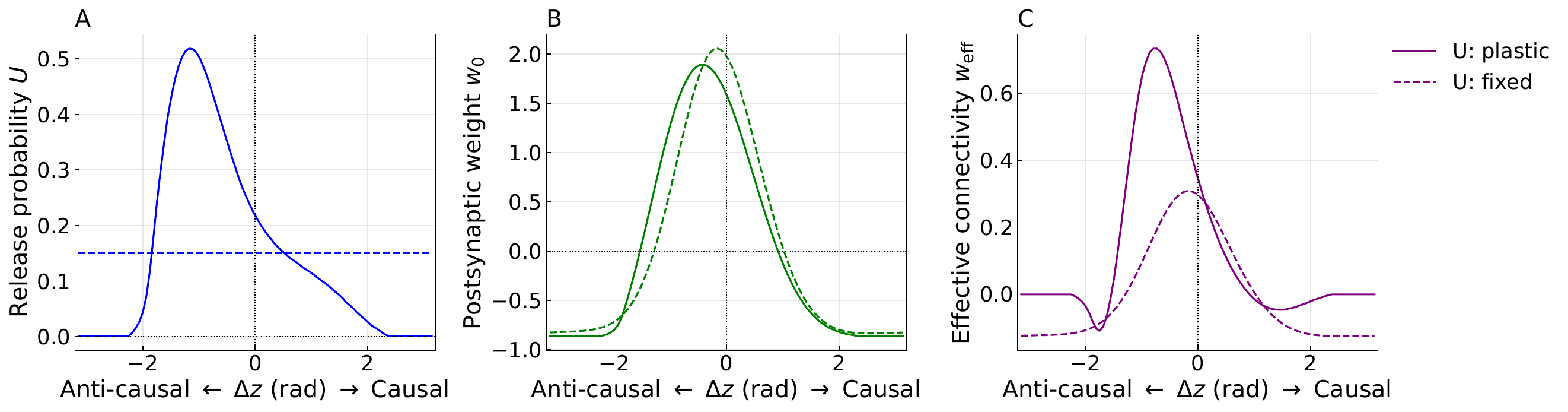}
    \caption[Optimal synaptic profiles.]{
    \textbf{Optimal synaptic profiles.} 
    \textbf{A.} Optimal release probability $U(\Delta z)$. 
    \textbf{B.} Optimal postsynaptic weight $w_0(\Delta z)$.
    \textbf{C.} Optimal effective weight $w_\text{eff}(\Delta z) = w_0(\Delta z) U(\Delta z)$.
    Solid: optimization with plastic $U(\Delta z)$.
    Dashed: optimization with a non-plastic $U$ fixed at 0.15.
    Optimal $U(\Delta z)$ is derived under the constraint $\langle U(\Delta z)\rangle = 0.15$.
    Other parameters are identical to Figure~\ref{fig:pre-post_factors}.}
    \label{fig:optimal-weights}
\end{figure*}

As predicted in Sec.~\ref{subsec:hebbian-form}, the resulting optimal $U(\Delta z)$ (Figure~\ref{fig:optimal-weights}A) is strongly \emph{temporally asymmetric}:
release probability is maximal when the presynaptic phase \emph{lags} the postsynaptic phase ($\Delta z < 0$), and remains low when presynaptic activity leads ($\Delta z > 0$).
Thus, the weak-coupling optimum predicts an \emph{anti-causal} bias in the learned connectivity.
Optimizing $w_0(\Delta z)$ with this optimized $U(\Delta z)$ produces a similarly asymmetric $w_0(\Delta z)$ (Figure~\ref{fig:optimal-weights}B,C),
whereas holding $U(\Delta z)$ fixed substantially attenuates the asymmetry (Figure~\ref{fig:optimal-weights}B,C),
highlighting that plasticity of release probability is essential for expressing the full temporal bias.

\begin{figure*}[!htbp]
    \centering
    \includegraphics[width=0.8\linewidth]{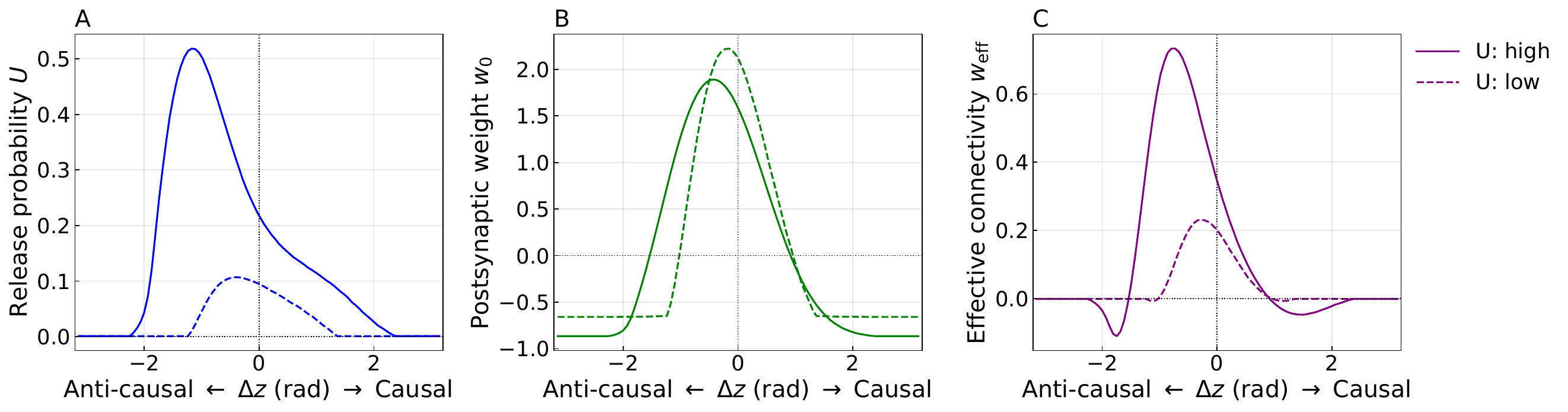}
    \caption[Low release probabilities attenuate temporal asymmetry.]{
    \textbf{Low release probabilities attenuate temporal asymmetry.} 
    Comparison of optimal $w(\Delta z)$ for different constraints on the mean release probability.
    Solid: $\langle U(\Delta z) \rangle= 0.15$.
    Dashed: $\langle U(\Delta z) \rangle = 0.03$.
    When release probability is constrained to low values, temporal asymmetry in the learning rule is substantially reduced, approaching a nearly symmetric profile.
    Other parameters are identical to Figure~\ref{fig:pre-post_factors}.}
    \label{fig:optimal-weights-awake-sleep}
\end{figure*}

The degree of temporal asymmetry depends on
constraints on release probability and on the strength of external inputs
(Figure~\ref{fig:optimal-weights-awake-sleep}; see also
Figs.~\ref{fig:optimize-U-supple-amplitudes} and ~\ref{fig:optimize-U-supple-constraints}).
Allowing higher release probabilities or stronger inputs consistently enhances
the anti-causal bias of both $U(\Delta z)$ and $w(\Delta z)$, whereas low release
probabilities lead to nearly symmetric potentiation and depression profiles.

These findings indicate that the effective learning rule is modulated by global constraints
on synaptic reliability and input drive. In particular, regimes with low release probability suppress
anti-causal biases and recover almost symmetric weight updates. In this low-$U$ limit, short-term synaptic dynamics
are effectively muted, so that updates closely resemble the case without STP, whereas high release probability
amplifies anti-causal associations. This constraint-dependent modulation provides a plausible mechanism for
state-dependent changes in synaptic learning rules, which we further relate to sleep--wake transitions in the
Discussion.

\subsubsection{Associative STP promotes informative anti-causal structure in recurrent networks}
\begin{figure*}[htbp]
    \centering    \includegraphics[width=0.8\linewidth]{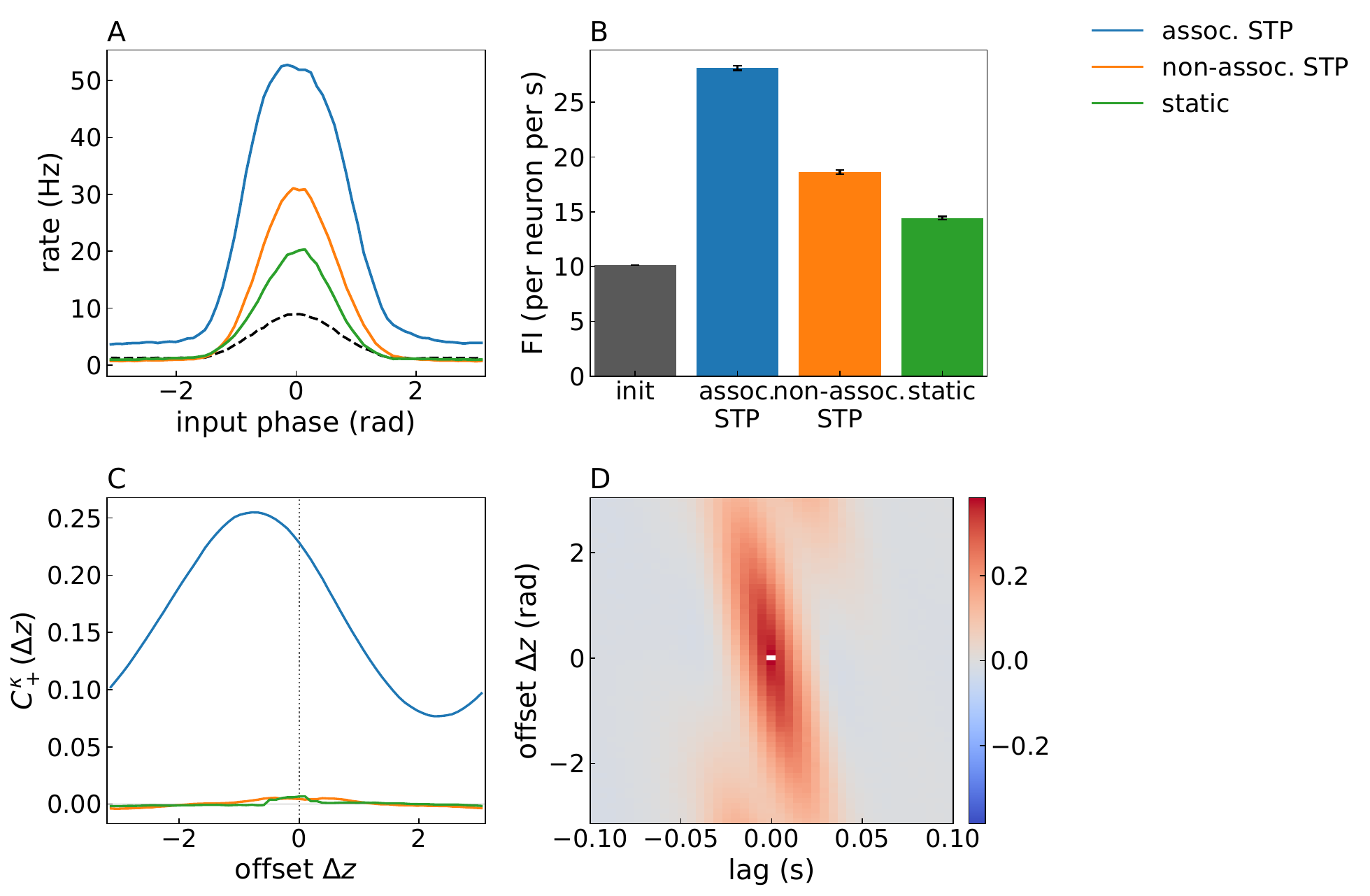}
\caption[Condition-dependent changes in evoked activity, Fisher information, and spontaneous correlations]{
\textbf{Condition-dependent changes in evoked activity, Fisher information, and spontaneous correlations.}
\textbf{A.} Phase-aligned population rate $r(\phi)$ during stimulus presentation, plotted against
$\phi=\omega t-z$.
Curves show the initial network (init; dashed black) and the networks after Fisher-information optimization under
associative STP (blue), non-associative STP (orange), and static synapses (green).
\textbf{B.} Fisher information before optimization (init; gray) and after optimization for each condition
(mean $\pm$ SE), normalized per neuron per second.
\textbf{C.} EPSP-kernel-weighted positive-lag correlation profile
$C_+^\kappa(\Delta z)$ during spontaneous activity for the optimized networks.
Profiles are computed from trial-shuffle-corrected spike-train cross-correlograms as
$C_+^\kappa(\Delta z)=
\sum_{0<\tau\le 5\tau_s} e^{-\tau/\tau_s} C(\Delta z,\tau)/
\sum_{0<\tau\le 5\tau_s} e^{-\tau/\tau_s}$.
\textbf{D.} Trial-shuffle-corrected spontaneous cross-correlogram
$C(\Delta z,\tau)$ for the optimized associative-STP network.
The zero-lag autocorrelation bin at $\Delta z=0$ is omitted.
During spontaneous activity (\textbf{C,D}), the traveling-wave drive is removed and the network receives a
uniform background input $h_{\mathrm{bg}}=0.5$.
Learning is performed under resource constraints: 
$w_0(\Delta z)$ is balanced (zero mean) and has a bounded normalized
$L_2$ norm, $\left[(2\pi)^{-1}\int w_0(\Delta z)^2\,d\Delta z\right]^{1/2}\le C=1.0$.
In the non-associative condition, we fix $U(\Delta z)=0.15$.
Simulations use $N=64$ neurons with a sigmoid nonlinearity
$g(u)=g_M[1+\exp(-\beta(u-u_c))]^{-1}$ ($g_M=500~\mathrm{Hz}$, $\beta=2.0$, $u_c=3.0$),
traveling-wave input amplitude $A=1.0$, angular frequency $\omega=2\pi$, and encoding parameter
$\theta_c=\pi/2$.
}
   \label{fig:evoked-activity}
\end{figure*}

The weak-coupling analysis above provides intuition for how associative STP biases
learning toward anti-causal associations.
We next test whether these predictions persist in a fully recurrent spiking
network by optimizing Fisher information using the simulation-based
gradient estimator (Sec.~\ref{sec:derive-learning-rule}) in this traveling-wave setting.
Because Fisher information \eqref{eq:FI-functional} can be increased trivially by raising firing rates,
we impose the same resource constraints as in Sec.~\ref{subsec:weak-coupling-prediction}: $w_0(\Delta z)$ is
constrained to be balanced (zero mean) and to have bounded $L_2$ norm (synaptic cost),
and the mean release probability is fixed during learning ($\langle U(\Delta z)\rangle = 0.15$).
Although these constraints are set empirically, we note that the learned stimulus-evoked firing rates (Figure~\ref{fig:evoked-activity}A) are broadly consistent with cortical responses to optimal stimuli reported experimentally \cite{gao2010Parallel,hawken2020Functional}.

We compare three synaptic models:
(i) \emph{associative STP}, in which both the postsynaptic weight profile
$w_0(\Delta z)$ and the release-probability profile $U(\Delta z)$ are optimized;
(ii) \emph{non-associative STP}, in which synapses exhibit Tsodyks--Markram
dynamics but $U(\Delta z)$ is held fixed and only $w_0(\Delta z)$ is learned; and
(iii) \emph{static} synapses without STP dynamics.
For reference, we also report the initial network before optimization (init).

Reflecting the anti-causal preference of the learning rule predicted in the preceding sections, learning with associative STP strengthens the latter part of the response within each stimulus epoch compared to static synapses (Figure~\ref{fig:evoked-activity}A, Fig.~\ref{fig:asymmetry-metrics}).
Furthermore, optimization improves Fisher information compared with the initial network across all synaptic models, 
with associative STP yielding the largest improvement, followed by non-associative STP and then static synapses (Figure~\ref{fig:evoked-activity}B). 

To relate these activity changes to circuit interactions, we computed pairwise
cross-correlograms during both stimulus-evoked and spontaneous activity as a
proxy for effective coupling (Figure~\ref{fig:evoked-activity}C,D; Fig.~\ref{fig:cross-correlograms}).
Associative STP yields clear anti-causal (``backward'') correlations during both evoked and spontaneous
activity (Figure~\ref{fig:evoked-activity}C,D; Figure~\ref{fig:cross-correlograms}A,D), whereas non-associative STP and static synapses show weaker
and less consistent temporal-correlation structure (Fig.~\ref{fig:cross-correlograms}B,C,E,F).

\subsubsection{STP supports reverse replay during spontaneous activity}

The results above indicate that learning with STP dynamics, particularly associative STP, can strengthen
effective coupling in the direction opposite to the experienced stimulus propagation.
Such ``backward'' connectivity motifs have been proposed as a circuit mechanism
for hippocampal \emph{reverse replay} \cite{haga2018Recurrent}.
We therefore asked whether Fisher-information optimization in our model gives
rise to reverse replay when external drive is removed.

\begin{figure*}[htbp]
    \centering
    \includegraphics[width=0.8\linewidth]{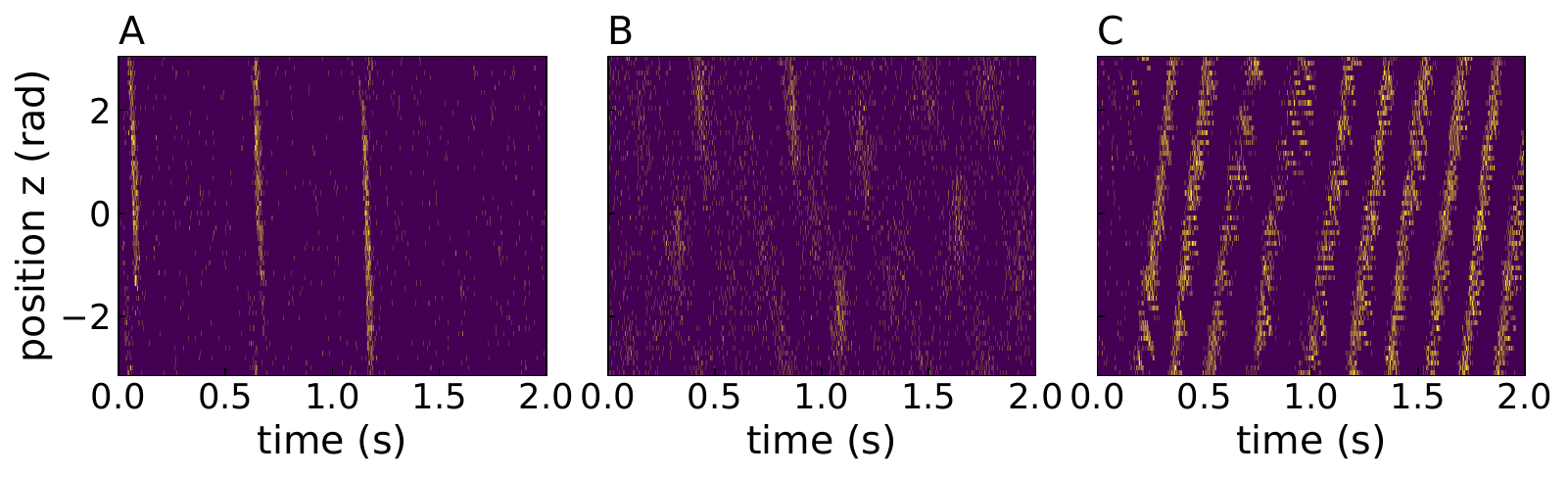}
    \caption[Emergence of reverse replay]{
    \textbf{Emergence of reverse replay.}
    Spontaneous activity after learning to optimize Fisher information (Figure~\ref{fig:evoked-activity}).
		    Reverse replay is observed in associative STP (\textbf{A}) and
		    non-associative STP (\textbf{B}), whereas static synapses show forward replay
		    (\textbf{C}).
		    Either type of replay is temporally compressed compared to the original stimulus.
		    During spontaneous activity, the traveling-wave drive is removed and the network receives a uniform
		    background input: $h_{\mathrm{bg}}=0.5$ (\textbf{A}), $h_{\mathrm{bg}}=1.5$ (\textbf{B}), and
		    $h_{\mathrm{bg}}=1.0$ (\textbf{C}).
		    Other parameters are identical to Figure~\ref{fig:evoked-activity}.
		    }
		   \label{fig:reverse-replay}
\end{figure*}

After learning, we simulated the network dynamics in the absence of stimulus
drive and examined the resulting spontaneous activity patterns.
With associative STP, spontaneous sequential events robustly propagated opposite to the original stimulus
direction, consistent with reverse replay (Figure~\ref{fig:reverse-replay}A).
Non-associative STP can also generate reverse replay, but requires a stronger background drive, and the resulting
sequences are less reliable (Figure~\ref{fig:reverse-replay}B).
In contrast, networks with static synapses require strong background input to generate replay-like events, which
propagate forward (Figure~\ref{fig:reverse-replay}C).

\section{Discussion}
We developed a Fisher-information-based theory for learning in synapses with
Tsodyks--Markram short-term dynamics, allowing both postsynaptic
strength and presynaptic release probability to be optimized.
The main consequence of introducing short-term depression into the learning
rule is that the effective presynaptic component becomes transient and
onset-sensitive.
Thus, for sequentially activated populations, maximizing encoded stimulus
information preferentially strengthens anti-causal connections, in which
presynaptic activity follows postsynaptic activity.
This bias is especially pronounced when release probability is learned
associatively, because increasing release probability both amplifies immediate
transmission and accelerates subsequent depletion.

In recurrent networks driven by traveling-wave input, this temporal bias
translated into more informative stimulus-evoked representations.
Under fixed resource constraints, all optimized networks increased Fisher
information relative to the initial network, but learning both $w_0$ and $U$
produced the largest gain, followed by non-associative STP and then static
synapses.
Associative STP also consistently enhanced backward effective
correlations during both evoked and spontaneous activity.
After removal of the traveling-wave drive, these learned interactions supported
reverse replay; non-associative STP could also generate reverse replay, but only
with stronger background drive and less reliable sequences, whereas static
synapses generated forward replay under comparable spontaneous conditions.

These results suggest that associative STP is not simply a faster version of
long-term potentiation.
Classical LTP modifies a relatively static postsynaptic efficacy, whereas
associative STP changes a presynaptic parameter that controls how synaptic
efficacy evolves during ongoing spike trains.
The same pre--post coincidence can therefore have different computational
effects depending on whether it changes $w_0$ or $U$: learning $U$ reshapes a
dynamic filter, making synaptic transmission sensitive to the timing and
frequency content of presynaptic activity.
This provides a compact normative interpretation of associative presynaptic
plasticity reported experimentally \cite{ucar2021Mechanical,kasai2023Mechanical}:
it can selectively amplify temporally informative components of an input stream
rather than merely increasing average coupling strength.

Our findings also connect Fisher-information optimization to circuit mechanisms
for replay.
Backward connectivity has been proposed as a substrate for hippocampal reverse
replay \cite{haga2018Recurrent}; here, such connectivity emerges from an
information-maximizing objective in the presence of short-term synaptic
dynamics.
Moreover, the strength of the anti-causal bias depends on the global constraint
on release probability.
Because synaptic profiles, such as presynaptic release probability and active-zone composition, are
state-dependent and can be modulated across sleep and wakefulness
\cite{hasselmo2006Role,sara2009Locus,huang2020Presynaptic,wu2025Presynaptic},
this resource dependence suggests a possible mechanism for regulating replay
directionality across behavioral states.
In this view, generous release-probability budgets favor reverse or
bidirectional replay, potentially useful for rapid credit assignment after
experience, whereas tighter budgets attenuate the reverse component and may
favor forward replay during memory stabilization
\cite{foster2006Reverse,diba2007Forward,ambrose2016Reverse,buzsaki2015Hippocampal,olafsdottir2018Role}.

The present theory has several limitations.
We focused on short-term depression and treated facilitation as negligible,
which is appropriate for isolating the interaction between associative learning
and slowly recovering synaptic resources but should be extended to
facilitation-dominated synapses.
We also used conditionally Poisson spiking, did not model explicit downstream
decoders, and did not enforce Dale's law or cell-type-specific inhibitory
dynamics.
These simplifications make the Fisher-information gradients analytically and
computationally tractable, but future work should test whether the same
principles hold in models with refractoriness, structured inhibition, explicit
decoding circuits, and alternative objectives such as mutual information or
predictive coding.
Finally, because STP parameters remain difficult to measure directly in large
populations \textit{in vivo}, statistical inference of synaptic dynamics from
population recordings \cite{ghanbari2017Estimating,ren2022Predictable} may
provide an experimental route for testing whether release-probability
constraints and associative STP covary with replay direction and behavioral
state.

In summary, optimizing information transmission in dynamic synapses yields
learning rules that differ qualitatively from static-weight plasticity.
By combining correlation detection with short-term synaptic filtering,
associative modulation of release probability can reshape neural
representations, enhance encoded stimulus information, and create recurrent
connectivity capable of supporting reverse replay.

\section*{CODE AVAILABILITY}
The source code used to generate the simulation results and figures is available at \url{https://github.com/genkinanodesu/AssociativeSTP}.

\section*{ACKNOWLEDGMENTS}
This work was funded by JSPS KAKENHI Grant Number 23KJ0666, JST CREST JPMJCR23N2, JSPS KAKENHI 25K24466, RIKEN Center for Brain Science, and RIKEN TRIP initiative (RIKEN
Quantum). 
We thank K. Yoshida, K. Inokuchi, T. Haga, and all members of the Toyoizumi lab for discussions.


\clearpage
\appendix 
\numberwithin{equation}{section}
\renewcommand{\thefigure}{S\arabic{figure}}
\renewcommand{\theHfigure}{S\arabic{figure}}
\setcounter{figure}{0}

\section{Exact gradient formula}
\label{appx:general-gradient}

This appendix derives the exact gradient expression
\eqref{eq:general-grad-score-identity} and the explicit forms
\eqref{eq:pathwise-term}--\eqref{eq:score-term} used in
Section~\ref{subsec:grad-computation}.

\subsection{Point-process likelihood and the score-function identity}
\label{appx:score-identity}

Let $N_i(t)$ denote the counting process of neuron $i$ and $dN_i(t)$ its increment.
Under our conditional independence assumption, the joint likelihood of the population spike history
$X(T)$ factorizes across neurons given the conditional intensities $\rho_i(t)=g(u_i(t))$:
\begin{multline}    
\label{eq:loglik-pointprocess}
\log P(X(T)\mid Z) \\
=
\sum_{k=1}^N
\left[
\int_0^T \log \rho_k(t)\, dN_k(t)
-
\int_0^T \rho_k(t)\,dt
\right].
\end{multline}
(Equivalently, $\int_0^T \varphi(t)\, dN_k(t)=\sum_{f_k}\varphi(t_k^{f_k})$ for any test function $\varphi$.)

For any scalar objective written as an expectation over spike histories,
\begin{align}
J(Z)=\left\langle \mathcal{J}[X,Z]\right\rangle_{X(T)}
=\int \mathcal{J}[X,Z]\,P(X(T)\mid Z)\,dX,
\end{align}
differentiation under the integral gives the standard score-function (likelihood-ratio) identity
\begin{multline}    
\label{eq:score-identity-general}
\frac{\partial J}{\partial Z} 
=
\left\langle
\frac{\partial \mathcal{J}[X,Z]}{\partial Z} 
+
\mathcal{J}[X,Z]\,
\frac{\partial}{\partial Z}\log P(X(T)\mid Z)
\right\rangle_{X(T)}.
\end{multline}
Applying \eqref{eq:score-identity-general} to $Z=Z_{ij}$ yields
\eqref{eq:general-grad-score-identity}.

Differentiating \eqref{eq:loglik-pointprocess} gives
\begin{multline}
\label{eq:score-general-pointprocess}
\frac{\partial}{\partial Z}\log P(X(T)\mid Z)
\\ 
= \sum_{k=1}^N
\int_0^T \left[dN_k(t)-\rho_k(t)\,dt\right]\,
\frac{\partial}{\partial Z}\log \rho_k(t).
\end{multline}
In our model $\rho_k(t)=g(u_k(t))$, hence
\begin{align}
\label{eq:dlogrho-du}
\frac{\partial}{\partial Z}\log \rho_k(t)
=
\frac{g_k'(t)}{g_k(t)}\,\frac{\partial u_k(t)}{\partial Z},
\qquad
g_k'(t):=\left.\frac{dg}{du}\right|_{u=u_k(t)}.
\end{align}
For the synapse-specific parameter $Z_{ij}\in\{w_{ij}^0,U_{ij}\}$, the membrane potential $u_k(t)$
depends on $Z_{ij}$ \emph{directly} only when $k=i$. Therefore
\eqref{eq:score-general-pointprocess}--\eqref{eq:dlogrho-du} reduce to
\begin{multline}
\label{eq:score-specific}
\frac{\partial}{\partial Z_{ij}}\log P(X(T)\mid Z) \\
=\int_0^T \left[dN_i(t)-\rho_i(t)\,dt\right]\,
\frac{g_i'(t)}{g_i(t)}\,
\underbrace{\frac{\partial u_i(t)}{\partial Z_{ij}}}_{=:~e_{ij}^{Z}(t)},
\end{multline}
which is \eqref{eq:score-term} in the main text.

\subsection{Pathwise derivative of the Fisher-information functional}
\label{appx:pathwise-derivative}

We next compute $\partial \mathcal{J}[X]/\partial Z_{ij}$ for the Fisher-information functional
\eqref{eq:FI-functional}. Define
\begin{align}
A_k(t) := h_k'(t,\theta)\frac{g_k'(t)}{g_k(t)},
\qquad
\rho_k(t)=g_k(t):=g(u_k(t)).
\end{align}
Then \eqref{eq:FI-functional} is
\begin{align}
\mathcal{J}[X]
=
\int_0^T dt \sum_{k=1}^N A_k(t)^2\,\rho_k(t).
\end{align}
For a fixed spike history $X(T)$, the dependence on $Z_{ij}$ enters only through the membrane potentials $u_k(t)$,
and hence through $g_k(t),g_k'(t),g_k''(t)$. Differentiating with the chain rule gives
\begin{align}
\frac{\partial \mathcal{J}[X]}{\partial Z_{ij}}
&=
\int_0^T dt\;
\frac{\partial}{\partial u_i}
\left(
h_i'(t,\theta)^2\,
\frac{g_i'(t)^2}{g_i(t)}
\right)\,
\frac{\partial u_i(t)}{\partial Z_{ij}}.
\end{align}
A short calculation yields
\begin{align}
\label{eq:ddu_gprime2_over_g}
\frac{\partial}{\partial u}
\left(\frac{g'(u)^2}{g(u)}\right)
=
\frac{g'(u)^2}{g(u)}
\left(
\frac{2g''(u)}{g'(u)}-\frac{g'(u)}{g(u)}
\right).
\end{align}
Using $\rho_i(t)=g_i(t)$ and defining $\eta_i(t)$ by \eqref{eq:eta-general}, we obtain
\begin{align}
\frac{\partial \mathcal{J}[X]}{\partial Z_{ij}}
&=
\int_0^T dt\;
\rho_i(t)\,\eta_i(t)\,
\underbrace{\frac{\partial u_i(t)}{\partial Z_{ij}}}_{=:~e_{ij}^{Z}(t)}.
\end{align}
This establishes \eqref{eq:pathwise-term}.

\subsection{A note on unbiased estimation and variance reduction}
\label{appx:mc-estimator}

Combining \eqref{eq:general-grad-score-identity}, \eqref{eq:score-specific}, and the pathwise term yields an
unbiased estimator from simulated trajectories:
\begin{multline}
\label{eq:mc-estimator}
    \widehat{\frac{\partial J}{\partial Z_{ij}}}
=
\frac{1}{M}\sum_{m=1}^M
\Bigg[
\int_0^T dt\; \rho_i^{(m)}(t)\eta_i^{(m)}(t)e_{ij}^{Z,(m)}(t) \\
+
\mathcal{J}[X^{(m)}]
\int_0^T
\left(dN_i^{(m)}(t)-\rho_i^{(m)}(t)\,dt\right)
\frac{g_i^{\prime,(m)}(t)}{g_i^{(m)}(t)}e_{ij}^{Z,(m)}(t)
\Bigg].
\end{multline}
As is common for score-function estimators, the second term can have high variance. A standard control variate is to
replace $\mathcal{J}[X^{(m)}]$ by $\mathcal{J}[X^{(m)}]-b$ with a constant $b$ (e.g., an online running mean),
which preserves unbiasedness because $\langle \partial_Z \log P\rangle=0$:
\begin{multline}    
\left\langle
b\,\frac{\partial}{\partial Z}\log P(X(T)\mid Z)
\right\rangle_{X(T)} \\= b\,\frac{\partial}{\partial Z}\int P(X(T)\mid Z)\,dX = 0.
\end{multline}

\section{Eligibility traces for STD synapses}
\label{appx:eligibility-traces}

This appendix provides explicit online update equations for the eligibility trace
$e_{ij}^Z(t)=\partial u_i(t)/\partial Z_{ij}$ for $Z_{ij}\in\{w_{ij}^0,U_{ij}\}$, under the Tsodyks--Markram
STD-only dynamics used in the main text.

\subsection{Eligibility as a filtered, parameter-weighted presynaptic spike train}

Write the presynaptic spike train as $x_j(t)=\sum_{f}\delta(t-t_j^f)$ and define the \emph{weighted} spike train
\begin{align}
y_{ij}(t):=\sum_f w_{ij}(t_j^f)\,\delta(t-t_j^f)=w_{ij}(t^-)\,x_j(t),
\end{align}
where $t^-$ indicates evaluation immediately before a spike at time $t$.
Then the synaptic contribution from neuron $j$ to $i$ can be written as a convolution
\begin{equation}
\begin{aligned}
u_i(t)=h_i(t,\theta)+\sum_{k=1}^N (\epsilon * y_{ik})(t), \\
(\epsilon * y)(t):=\int_0^t \epsilon(t-s)\,y(s)\,ds.
\end{aligned}
\end{equation}
Differentiating w.r.t.\ $Z_{ij}$ gives
\begin{equation}
\begin{aligned}
\label{eq:eligibility-convolution}
e_{ij}^Z(t)
=
\frac{\partial u_i(t)}{\partial Z_{ij}}
=
(\epsilon * \psi_{ij}^Z)(t),
\\
\psi_{ij}^Z(t):=\frac{\partial y_{ij}(t)}{\partial Z_{ij}}
=\frac{\partial w_{ij}(t^-)}{\partial Z_{ij}}\,x_j(t).
\end{aligned}
\end{equation}

\subsection{Exponential EPSP kernel: ODE / event-driven update}

For $\epsilon(t)=e^{-t/\tau_m}\Theta(t)$, the convolution representation \eqref{eq:eligibility-convolution}
is equivalent to the linear ODE
\begin{align}
\label{eq:eligibility-ode}
\dot e_{ij}^Z(t)
=
-\frac{1}{\tau_m}e_{ij}^Z(t)
+
\frac{\partial w_{ij}(t^-)}{\partial Z_{ij}}\,x_j(t).
\end{align}
Hence $e_{ij}^Z(t)$ decays exponentially between presynaptic spikes and exhibits jumps at spike times:
\begin{align}
\begin{cases}
\text{between spikes:}\quad &e_{ij}^Z(t+\Delta t)=e_{ij}^Z(t)\,e^{-\Delta t/\tau_m},
\\
\text{at }t=t_j^f:\quad &e_{ij}^Z(t^+)=e_{ij}^Z(t^-)+\frac{\partial w_{ij}(t^-)}{\partial Z_{ij}}.
\end{cases}
\end{align}

\subsection{STD dynamics and parameter sensitivities}

Under the STD-only assumption of the main text, the synaptic efficacy is
\begin{align}
w_{ij}(t)=w_{ij}^0\,U_{ij}\,d_{ij}(t),
\qquad d_{ij}(t)\in[0,1],
\end{align}
and the depression variable evolves as
\begin{align}
\label{eq:std-xj-form}
\dot d_{ij}(t)=\frac{1-d_{ij}(t)}{\tau_d}-U_{ij}\,d_{ij}(t^-)\,x_j(t).
\end{align}
Define the parameter sensitivity of $d_{ij}$ by
\begin{align}
s_{ij}^Z(t):=\frac{\partial d_{ij}(t)}{\partial Z_{ij}}.
\end{align}
For $Z_{ij}=w_{ij}^0$, $d_{ij}(t)$ does not depend on $w_{ij}^0$, hence
\begin{align}
\label{eq:sd-w0}
s_{ij}^{w_0}(t)\equiv 0.
\end{align}
For $Z_{ij}=U_{ij}$, differentiating \eqref{eq:std-xj-form} yields the distributional ODE
\begin{align}
\label{eq:sd-U-ode}
\dot s_{ij}^{U}(t)
=
-\frac{1}{\tau_d}s_{ij}^{U}(t)
-
\left[d_{ij}(t^-)+U_{ij}\,s_{ij}^{U}(t^-)\right]x_j(t),
\end{align}
which corresponds to the event-driven update
\begin{align}
\begin{cases}
\text{between spikes:}\quad &s_{ij}^U(t+\Delta t)=s_{ij}^U(t)\,e^{-\Delta t/\tau_d},\\
\text{at }t=t_j^f:\quad &s_{ij}^U(t^+)= (1-U_{ij})\,s_{ij}^U(t^-)-d_{ij}(t^-).
\end{cases}
\end{align}

Finally, the parameter derivatives of the synaptic efficacy (evaluated at spike arrival times) are
\begin{equation}
\label{eq:dw-dZ}
\begin{aligned}
\frac{\partial w_{ij}(t^-)}{\partial w_{ij}^0}
&=
U_{ij}\,d_{ij}(t^-),\\
\frac{\partial w_{ij}(t^-)}{\partial U_{ij}}
&=
w_{ij}^0\left[d_{ij}(t^-)+U_{ij}\,s_{ij}^U(t^-)\right].
\end{aligned}    
\end{equation}
Substituting \eqref{eq:dw-dZ} into \eqref{eq:eligibility-ode} yields an explicit online computation of $e_{ij}^Z(t)$ from the simulated
presynaptic spikes and STP state variables.

\section{Weak-coupling reduction: factorization and closed STD sensitivity dynamics}
\label{appx:weak-coupling-compensation}

This appendix collects derivations that are specific to the weak-coupling baseline ($w=0$):
(i) why the score term vanishes at leading order,
(ii) how the Poisson compensation yields a factorized form,
and (iii) how the same weak-coupling closure implies closed dynamics for the mean synaptic efficacy
and the normalized sensitivity functions $f^{w_0}$ and $f^{U}$ used in the main text.

\subsection{Why the score term vanishes at \texorpdfstring{$w=0$}{w=0}}

At $w_{ij}^0\equiv 0$, the membrane potentials satisfy $u_i(t)=h_i(t,\theta)$ deterministically,
so $\rho_i(t)=\nu_i^0(t):=g(h_i(t,\theta))$ is non-random. Consequently, the Fisher-information functional
$\mathcal{J}[X]$ becomes deterministic (i.e., independent of $X$), and the score term in
\eqref{eq:general-grad-score-identity} vanishes:
\begin{align}
\begin{split}
\left\langle
\mathcal{J}[X]\,
\frac{\partial}{\partial Z_{ij}}\log P(X(T)\mid Z)
\right\rangle_{w=0} \\
=
\mathcal{J}_0\,
\left\langle
\frac{\partial}{\partial Z_{ij}}\log P(X(T)\mid Z)
\right\rangle_{w=0} \\
=
\mathcal{J}_0\,
\frac{\partial}{\partial Z_{ij}}
\int P(X(T)\mid Z)\,dX
=0.
\end{split}
\end{align}

\subsection{Compensation formula at \texorpdfstring{$w=0$}{w=0} and the factorized expectation}

At $w=0$, each neuron fires as an inhomogeneous Poisson process with deterministic intensity $\nu_i^0(t)$.
Let $\phi(t)$ be any predictable process (measurable w.r.t.\ the past spike history up to $t^-$).
Then the Doob--Meyer decomposition implies the compensation formula
\begin{align}
\label{eq:compensation}
\left\langle
\int_0^T \phi(t)\,dN_j(t)
\right\rangle_{w=0}
=
\left\langle
\int_0^T \phi(t)\,\nu_j^0(t)\,dt
\right\rangle_{w=0}.
\end{align}
Because $\nu_j^0(t)$ is deterministic at $w=0$, it can be taken outside the expectation, yielding
\begin{align}
\left\langle
\int_0^T \phi(t)\,dN_j(t)
\right\rangle_{w=0}
=
\int_0^T \nu_j^0(t)\,\left\langle \phi(t)\right\rangle_{w=0}\,dt.
\end{align}
Applying this identity to the eligibility representation
$e_{ij}^Z(t)=\int_0^t \epsilon(t-t')\,\partial_{Z_{ij}}w_{ij}(t'^-)\,dN_j(t')$
yields the factorized form used in \eqref{eq:FI-grad}:
\begin{align}
\left\langle e_{ij}^Z(t)\right\rangle_{w=0}
=
\int_0^t dt'\,\epsilon(t-t')\,\nu_j^0(t')\,
\frac{\partial}{\partial Z_{ij}}\left\langle w_{ij}(t')\right\rangle_{w=0}.
\end{align}
Substituting this into the pathwise term \eqref{eq:pathwise-term} at $w=0$ gives \eqref{eq:FI-grad}.

\subsection{Closed dynamics of the mean efficacy and normalized sensitivities}
\label{appx-subsec:fw0fU-derivation}

Here we derive the closed dynamics of the normalized sensitivity functions used in the main text.
This derivation is conceptually part of the weak-coupling reduction: it relies only on the baseline
Poisson statistics (as above) and does not introduce any additional approximation beyond $w\to 0$.

For notational simplicity, we omit synaptic indices and write $w(t)$, $w^0$, $U$, and $\nu^0(t)$.

\subsubsection{Mean dynamics under Poisson spiking}

Under STD-only Tsodyks--Markram dynamics, the effective synaptic strength at presynaptic spike times can be written as
\begin{equation}
    w(t) = w^0 U d(t),
\end{equation}
with the equivalent event-driven representation
\begin{equation}
    \dot{w}(t)
    = \frac{w^0 U - w(t)}{\tau_d}
      - U w(t^-) \delta\bigl(t - t^{\text{spike}}\bigr),
\end{equation}
where $t^{\text{spike}}$ denotes presynaptic spike times.

Define the conditional mean
\begin{equation}
    m(t) := \bigl\langle w(t) \bigr\rangle_{X(t)}.
\end{equation}
Under weak coupling, the presynaptic spike train is unaffected by $w^0$ and $U$ and can be treated as an inhomogeneous
Poisson process with deterministic rate $\nu^0(t)$. Then the standard identity holds:
\begin{equation}
    \bigl\langle w(t^-) \delta\bigl(t - t^{\text{spike}}\bigr) \bigr\rangle_{X(t)}
    = \nu^0(t)\, \bigl\langle w(t) \bigr\rangle_{X(t)}
    = \nu^0(t)\, m(t).
\end{equation}
Taking the ensemble average yields the closed ODE
\begin{equation}
    \dot{m}(t)
    = \frac{w^0 U - m(t)}{\tau_d}
      - U \nu^0(t)\, m(t).
    \label{appx-eq:m-dynamics}
\end{equation}

\subsubsection{Parameter sensitivities and normalized sensitivity functions}

Differentiate \eqref{appx-eq:m-dynamics} with respect to $w^0$ (noting $\nu^0(t)$ and $U$ do not depend on $w^0$):
\begin{align}
    \frac{\mathrm{d}}{\mathrm{d}t}
    \frac{\partial m(t)}{\partial w^0}
    &=
    \frac{U}{\tau_d}
    -
    \left(\frac{1}{\tau_d} + U \nu^0(t)\right)
    \frac{\partial m(t)}{\partial w^0}.
    \label{appx-eq:dm-dw0}
\end{align}
Similarly, differentiation with respect to $U$ gives
\begin{align}
    \frac{\mathrm{d}}{\mathrm{d}t}
    \frac{\partial m(t)}{\partial U}
    &=
    \frac{w^0}{\tau_d}
    -
    \left(\frac{1}{\tau_d} + U \nu^0(t)\right)
    \frac{\partial m(t)}{\partial U}
    -
    \nu^0(t)\, m(t).
    \label{appx-eq:dm-dU}
\end{align}
Because \eqref{appx-eq:m-dynamics} is linear in $m(t)$ and proportional to $w^0$ (given our initial conditions),
the solution satisfies $m(t)=w^0\,y(t;U)$ and hence
\begin{equation}
    \frac{\partial m(t)}{\partial w^0} = \frac{m(t)}{w^0}.
    \label{appx-eq:m-propto-w0}
\end{equation}

Now define the normalized sensitivity functions (main text Eq.~\eqref{eq:def-fw0-fU}):
\begin{align}
    f^{w_0}(t) &:= \frac{1}{U}\,\frac{\partial m(t)}{\partial w^0},
    &
    f^{U}(t) &:= \frac{1}{w^0}\,\frac{\partial m(t)}{\partial U}.
\end{align}
Using \eqref{appx-eq:dm-dw0} yields
\begin{align}
    \dot{f}^{w_0}(t)
    &=
    \frac{1}{\tau_d}
    -
    \left(\frac{1}{\tau_d} + U \nu^0(t)\right) f^{w_0}(t),
    \label{appx-eq:fw0-dynamics}
\end{align}
and using \eqref{appx-eq:dm-dU} together with \eqref{appx-eq:m-propto-w0} yields
\begin{align}
    \dot{f}^{U}(t)
    &=
    \frac{1}{\tau_d}
    -
    \left(\frac{1}{\tau_d} + U \nu^0(t)\right) f^{U}(t)
    -
    \nu^0(t)\,U f^{w_0}(t).
    \label{appx-eq:fU-dynamics}
\end{align}
These are exactly the dynamics stated in the main text (Eq.~\eqref{eq:fw0fU-dynamics}, with $\nu^0(t)$ denoted there as $\nu_j^0(t)$).

\section{Analysis of STD sensitivity functions and frequency-domain response}
\label{appx-sec:std-sensitivity-response}

This appendix analyzes the properties of the sensitivity functions $f^{w_0}(t)$ and $f^{U}(t)$ and the effective
presynaptic learning term $C^Z(t)=f^Z(t)\nu(t)$, starting from the closed weak-coupling dynamics derived in
Appendix~\ref{appx-subsec:fw0fU-derivation} (equivalently, main text Eq.~\eqref{eq:fw0fU-dynamics}).
For notational simplicity, we omit synaptic indices and write $\nu(t)$ for the deterministic presynaptic rate in the
weak-coupling baseline (i.e., $\nu(t)\equiv \nu^0(t)$ in Eq.~\eqref{eq:fw0fU-dynamics}).

\subsection{Illustrative step and sinusoidal responses}
\begin{figure}[htbp]
    \centering
    \includegraphics[width=\linewidth]{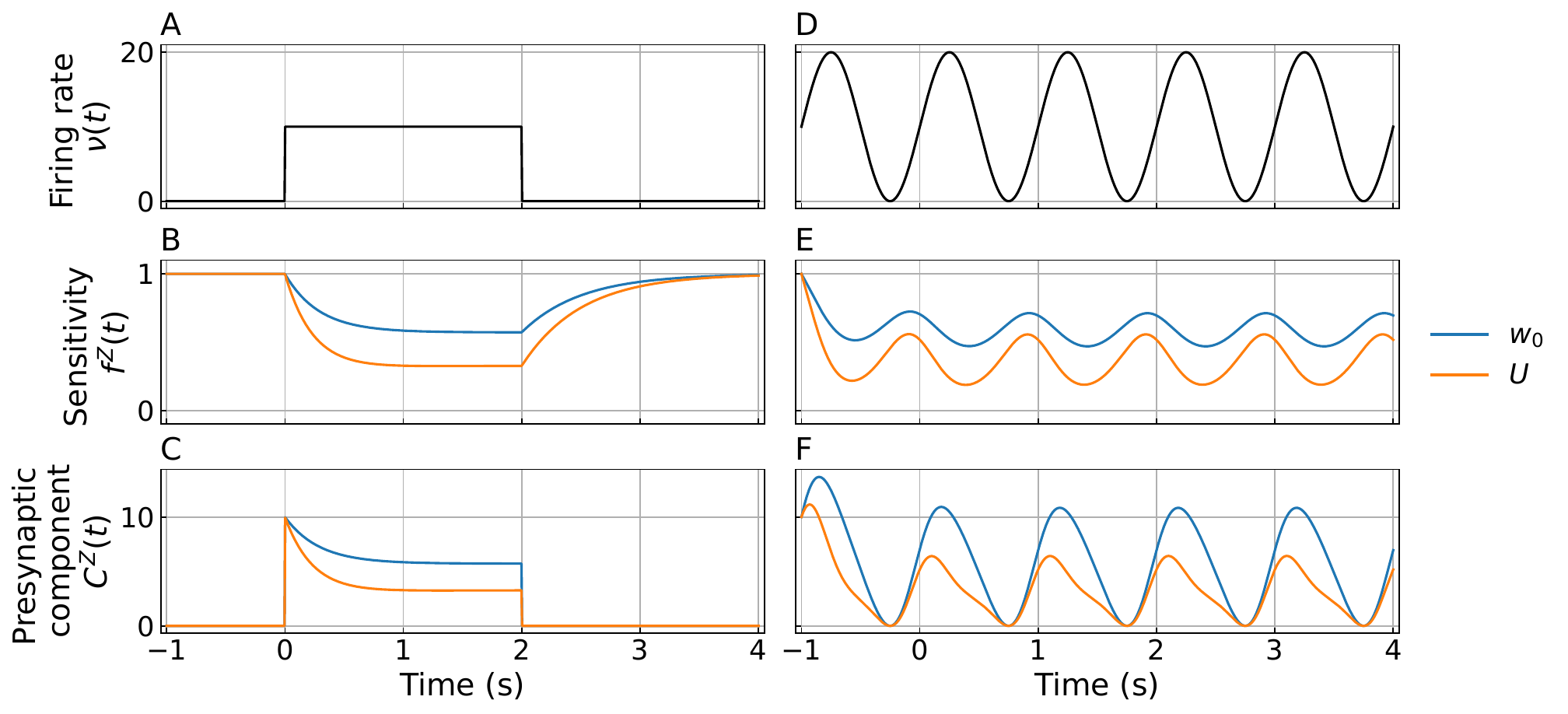}
    \caption[Response of STD sensitivity functions to step / sinusoidal inputs.]%
    {\textbf{Response of STD sensitivity functions to step / sinusoidal inputs.}
    \textbf{A--C}. Response to a step increase in presynaptic firing rate $\nu(t)$.
    \textbf{A}. Presynaptic firing rate $\nu(t)$.
    \textbf{B}. Sensitivity functions $f^{Z}(t)$. Note that $f^{U}$ (orange) decays more strongly than $f^{w_0}$
    (blue).
    \textbf{C}. Effective presynaptic learning term $C(t) = f(t)\nu(t)$.
    \textbf{D--F}. Response to sinusoidal modulation.
    The effective presynaptic term (F) peaks during the rising phase of the input (D), illustrating onset
    sensitivity of the learning rule.
    Parameters: $\tau_d = 0.5$~s, $U = 0.15$.}
    \label{fig:STP-component}
\end{figure}

Figure~\ref{fig:STP-component} provides time-domain intuition by showing the responses of $f^{w_0}(t)$, $f^{U}(t)$,
and the effective presynaptic term $C^Z(t)=f^Z(t)\nu(t)$ to a step increase and to a sinusoidal modulation of
$\nu(t)$. The following subsections quantify these effects.

\subsection{Response to constant input}
For a constant presynaptic firing rate $\nu$, the system of differential equations \eqref{eq:fw0fU-dynamics} yields the following steady-state solutions:
\begin{align}
\label{eq:fw0fU-convergence}
    f^{w_0}_* = \frac{1}{1 + \tau_d \nu U}, \quad 
    f^{U}_* = \frac{1}{(1 + \tau_d \nu U)^2}.
\end{align}
The transient responses from arbitrary initial conditions are given by:
\begin{align}
\label{eq:fw0fU-transient}
    f^{w_0}(t) &= f^{w_0}_* + \left[f^{w_0}(0) - f^{w_0}_*\right] e^{-\kappa t} \\
    f^{U}(t) &= f^{U}_* + \bigg[\left(f^{U}(0) - f^{U}_*\right) - \nu U \left(f^{w_0}(0) - f^{w_0}_*\right) t\bigg] e^{-\kappa t},
\end{align}
where $\kappa = \tau_d^{-1} + \nu U$ represents the effective decay rate.

\subsection{Linear response to sinusoidal modulation}
\label{appx-subsec:linear-response-to-sin-modulation}
We analyze the linear response of the sensitivity functions and of the effective presynaptic term $C^Z(t)=f^Z(t)\nu(t)$ to a weak sinusoidal modulation of the presynaptic firing rate. We consider a rate modulation of the form
\begin{equation}
    \nu(t) = \nu_0 + \delta \nu \cos(\omega t) = \nu_0 + \Re\left[ \hat{\nu} e^{i\omega t} \right],
\end{equation}
where $\nu_0$ is the baseline rate, $\hat{\nu}$ is the small amplitude of modulation ($|\hat{\nu}| \ll \nu_0$), and $\omega$ is the angular frequency.

\subsubsection{Linearization of the dynamics.}
We decompose the sensitivity functions into their steady-state values and small time-dependent fluctuations:
\begin{equation}
    f^{w_0}(t) = f^{w_0}_* + \delta f^{w_0}(t), \qquad 
    f^{U}(t) = f^{U}_* + \delta f^{U}(t).
\end{equation}
Substituting these into Eq.~\eqref{eq:fw0fU-dynamics} and retaining only first-order terms in $\delta f$ and $\delta \nu$, we obtain the linearized system:
\begin{align}
    \dot{\delta f}^{w_0}(t) &= -\kappa \delta f^{w_0}(t) - U f^{w_0}_* \delta \nu(t), \\
    \dot{\delta f}^{U}(t) &= -\kappa \delta f^{U}(t) - \nu_0 U \delta f^{w_0}(t) - U(f^{U}_* + f^{w_0}_*) \delta \nu(t),
\end{align}
where $\kappa = \tau_d^{-1} + \nu_0 U$ is the effective decay rate derived in the previous subsection.

Switching to the frequency domain with $\delta f^Z(t) = \Re[\hat{f}^Z e^{i\omega t}]$, the complex transfer functions $H_{f}^Z(\omega) = \hat{f}^Z / \hat{\nu}$ are obtained as:
\begin{align}
    H_{f}^{w_0}(\omega) &= -\frac{U f^{w_0}_*}{\kappa + i\omega}, \\
    H_{f}^{U}(\omega) &= -\frac{U(f^{U}_* + f^{w_0}_*)}{\kappa + i\omega} + \frac{\nu_0 U^2 f^{w_0}_*}{(\kappa + i\omega)^2}.
\end{align}
These transfer functions show that the sensitivity functions behave as (inverted) low-pass filtered versions of the rate modulation.

\begin{figure}[htbp]
    \centering
    \includegraphics[width=\linewidth]{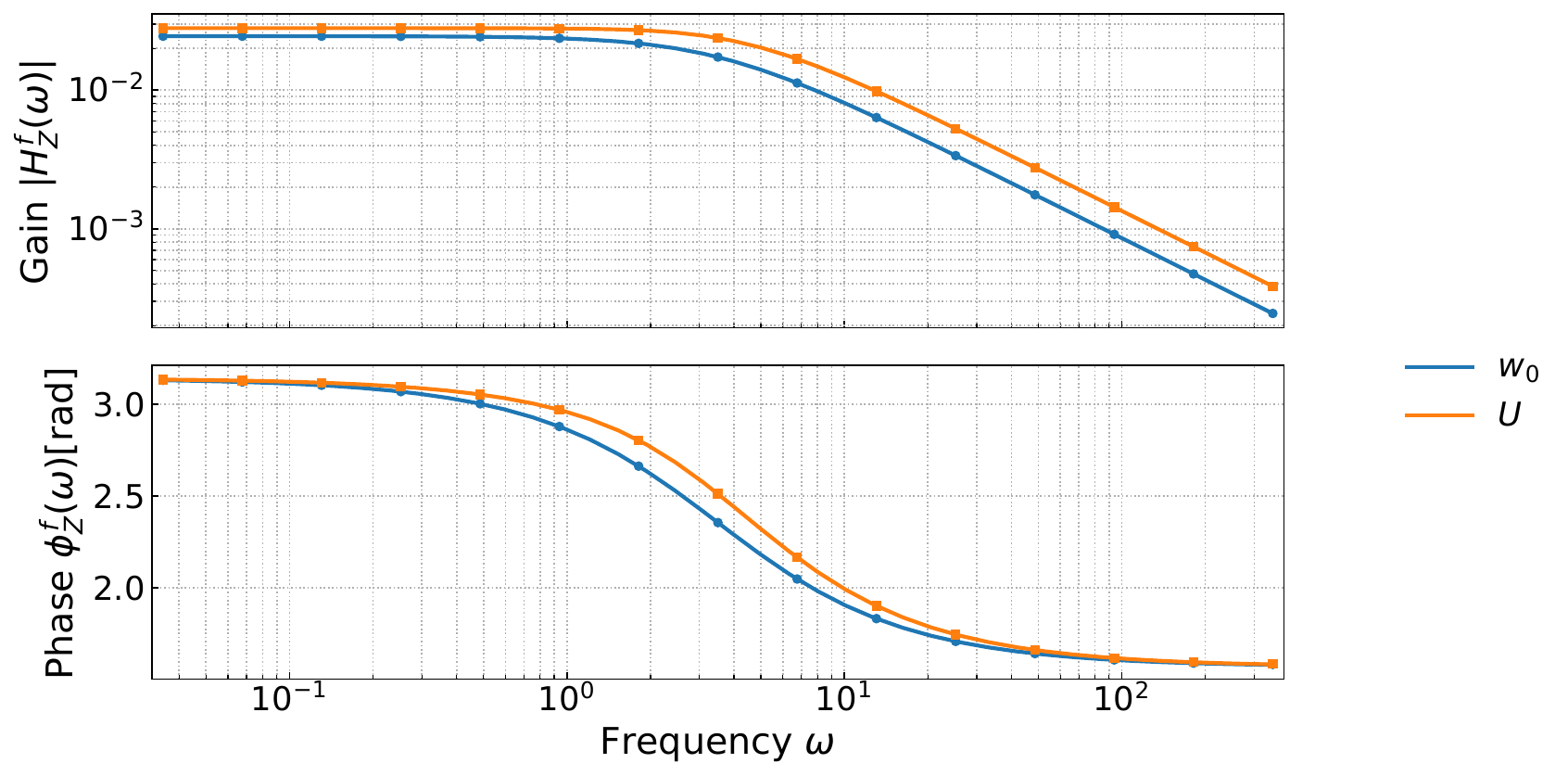}
    \caption[Frequency-domain response of sensitivity functions]
    {\textbf{Frequency-domain response of the raw sensitivity functions $f^{w_0}$ and $f^{U}$.}
    Amplitude gain (upper) and phase lag (lower) of $f$ (not the product $C = f\nu$). These functions behave as low-pass filters with different effective orders.
    Parameters: $\tau_d = 0.5~$s, $\nu_0 = 10~$Hz, $U = 0.15$.
    }
    \label{fig:STP-phase-response-f}
\end{figure}

For $f^{w_0}$, the amplitude gain and phase lag relative to the input modulation can be explicitly derived as:
\begin{align}
    |H_{f}^{w_0}(\omega)| &= \frac{U f^{w_0}_*}{\sqrt{\kappa^2 + \omega^2}}, \\
    \phi_{f}^{w_0}(\omega) &= \pi - \tan^{-1}\left(\frac{\omega}{\kappa}\right).
\end{align}
From these expressions, it is evident that the (principal-value) phase $\phi_{f}^{w_0}(\omega)$ decreases monotonically from $\pi$ to $\pi/2$ as $\omega$ increases from $0$ to $\infty$ (Figure \ref{fig:STP-phase-response-f}).

For $f^{U}$, although the explicit decomposition is algebraically more complex, the phase lag similarly exhibits a monotonic decrease from $\pi$ to $\pi/2$. Crucially, however, due to the contribution of the second-order pole term, $f^U$ responds more sluggishly than $f^{w_0}$. Consequently, in the intermediate frequency range where $\omega \sim \kappa$, the phase lag of $f^U$ is consistently larger than that of $f^{w_0}$ (Figure \ref{fig:STP-phase-response-f}).

\subsubsection{Effective presynaptic term and frequency response}
The synaptic learning rule \eqref{eq:FI-grad-Hebbform} depends on the effective presynaptic term $C^Z(t) = f^Z(t)\nu(t)$. Its linearization yields:
\begin{equation}
    \delta C^Z(t) \approx f^Z_* \delta \nu(t) + \nu_0 \delta f^Z(t).
\end{equation}
The corresponding frequency response function $H_{C}^Z(\omega) = \hat{C}^Z / \hat{\nu}$ is therefore given by the superposition of the direct rate modulation and the filtered sensitivity dynamics:
\begin{equation}
    H_{C}^Z(\omega) = f^Z_* + \nu_0 H_{f}^Z(\omega).
\end{equation}

\begin{figure}[t]
    \centering
    \includegraphics[width=\linewidth]{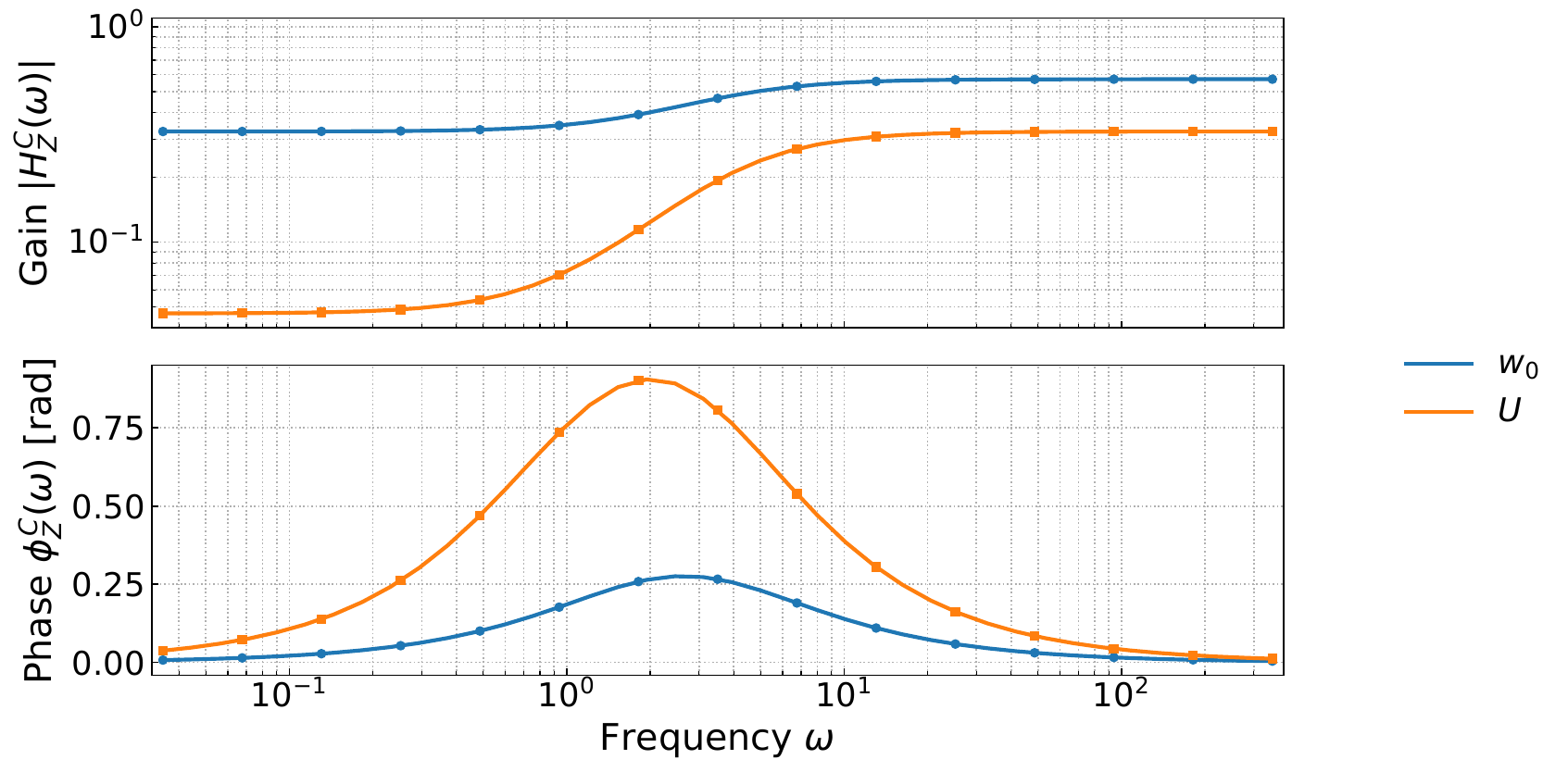}
    \caption[Frequency-domain response of the effective presynaptic term $C^{Z}(t)$.]%
    {\textbf{Frequency-domain response of the effective presynaptic term $C^{Z}(t)$.}
    Response to a modulated rate $\nu(t) = \nu_0 + \delta\nu \cos(\omega t)$.
    Upper: amplitude gain $|H_{C}^{Z}(\omega)|$. 
    Lower: phase $\phi_{C}^{Z}(\omega) = \arg H_{C}^{Z} (\omega)$. 
    Note the positive phase shift (lead) in the intermediate frequency range, indicating sensitivity to input
    onset. The lead is more pronounced for $U$ (orange) than for $w_0$ (blue).
    Lines show the predictions by linear response theory, and dots show results from numerical simulations.
    Parameters: $\nu_0 = 10$~Hz, $\tau_d = 0.5$~s, $U = 0.15$.
    }
    \label{fig:STP-phase-response-C}
\end{figure}

Figure~\ref{fig:STP-phase-response-C} shows the frequency response of the effective presynaptic term $C^Z(t)$ for a
representative operating point, comparing linear-response predictions with numerical simulations. Over an
intermediate frequency band, the phase becomes positive (a lead), indicating that the presynaptic contribution to
plasticity is biased toward the rising phase of the input. This lead is more pronounced for $Z=U$ than for
$Z=w_0$.

\subsubsection{Response characteristic for \texorpdfstring{$Z = w_0$}{Z = w0}.}
For the baseline weight parameter ($Z=w_0$), substituting $H_{f}^{w_0}(\omega)$ yields:
\begin{align}
    H_{C}^{w_0}(\omega)
    &= f^{w_0}_* \left( 1 - \frac{\nu_0 U}{\kappa + i\omega} \right)
      = \frac{1}{\tau_d \kappa} \frac{\tau_d^{-1} + i\omega}{\kappa + i\omega}.
\end{align}
This transfer function represents a lead--lag compensator. Since $\kappa > \tau_d^{-1}$, the phase is positive (leading) for all $\omega > 0$. Writing the phase as the difference between the arguments of the zero and the pole,
\begin{equation}
    \phi(\omega)
    = \arg H_{C}^{w_0}(\omega)
    = \arctan(\omega \tau_d) - \arctan\!\left(\frac{\omega}{\kappa}\right),
\end{equation}
its stationary points satisfy
\begin{equation}
    \frac{\mathrm{d}\phi}{\mathrm{d}\omega}
    = \frac{\tau_d^{-1}}{\tau_d^{-2} + \omega^2}
      - \frac{\kappa}{\kappa^2 + \omega^2}
    = 0.
\end{equation}
Solving this equation gives
\begin{equation}
    \omega^2 = \tau_d^{-1}\kappa,
\end{equation}
and hence the phase lead is maximized at the geometric mean of the pole and zero frequencies,
\begin{equation}
    \omega_{*, w_0} = \sqrt{\tau_d^{-1} \kappa}.
\end{equation}
Because $\phi(\omega) \to 0$ as $\omega \to 0$ and as $\omega \to \infty$, and $\phi(\omega) > 0$ for intermediate frequencies, this stationary point corresponds to the global maximum of the phase lead.

It is also instructive to examine the limiting cases of $H_{C}^{w_0}(\omega)$. In the quasi-static limit $\omega \to 0$, we obtain
\begin{equation}
    H_{C}^{w_0}(0)
    = \frac{1}{\tau_d \kappa} \frac{\tau_d^{-1}}{\kappa}
    = (\kappa \tau_d)^{-2}.
\end{equation}
Using $\kappa \tau_d = 1 + \tau_d \nu_0 U$, this low-frequency gain can be written as
\begin{equation}
    H_{C}^{w_0}(0)
    = \left.\frac{\mathrm{d}}{\mathrm{d}\nu}
      \bigl[ \nu f_*^{w_0}(\nu)\bigr]\right|_{\nu=\nu_0}
    = f_*^{w_0}(\nu_0) + \nu_0 \frac{\partial f_*^{w_0}}{\partial \nu_0},
\end{equation}
that is, it coincides with the derivative of the steady-state effective presynaptic term
$C_*^{w_0}(\nu) = \nu f_*^{w_0}(\nu)$ with respect to $\nu$, evaluated at $\nu_0$.
Thus, when the input varies sufficiently slowly, the response of $C^{w_0}(t)$ is
consistent with a quasi-static approximation in which the system adiabatically
tracks the steady-state relation between $\nu$ and $C^{w_0}$.

In the opposite limit $\omega \to \infty$, the transfer function reduces to
\begin{equation}
    \lim_{\omega \to \infty} H_{C}^{w_0}(\omega)
    = \frac{1}{\tau_d \kappa}
    = f^{w_0}_*.
\end{equation}
Here the oscillatory modulation is too fast for the sensitivity dynamics to
follow, so the fluctuation $\delta f^{w_0}(t)$ is effectively averaged out.
Consequently, only the instantaneous modulation of the firing rate contributes
and the gain of the effective presynaptic term is given by the constant
steady-state factor $f^{w_0}_*$.

\subsubsection{Response characteristic for \texorpdfstring{$Z = U$}{Z = U}.}
For the release probability parameter ($Z=U$), the derivation is more involved due to the coupled dynamics. Substituting $H_{f}^{U}(\omega)$ and simplifying the algebraic terms, we obtain:
\begin{align}
    H_{C}^U (\omega)
    = \frac{1}{(\tau_d \kappa)^2} \,
      \frac{\tau_d^{-1} + i\omega}{(\kappa + i\omega)^2}
      \left(-\tau_d \kappa^2 + 2 \kappa + i\omega\right).
\end{align}
To analyze the phase, it is convenient to introduce the dimensionless parameters
\begin{equation}
    r := \kappa \tau_d, \qquad x := \frac{\omega}{\kappa}.
\end{equation}
Using $\tau_d^{-1} = \kappa / r$ and $-\tau_d \kappa^2 + 2\kappa = \kappa(2-r)$, the transfer function can be rewritten as
\begin{equation}
    H_{C}^U(\omega)
    = \frac{1}{r^2} \,
      \frac{\bigl(\frac{1}{r} + i x\bigr)\bigl((2-r) + i x\bigr)}{(1 + i x)^2},
\end{equation}
where the positive prefactor $1/r^2$ does not affect the phase. Thus, the phase response $\phi_U(\omega) = \arg H_C^U(\omega)$ can be expressed as
\begin{equation}
    \phi_U(\omega)
    = \phi_U(x)
    = \arctan(r x) + \arctan\!\left(\frac{x}{2-r}\right)
      - 2 \arctan(x).
\end{equation}
The stationary points of the phase satisfy
\begin{equation}
    \frac{\mathrm{d}\phi_U}{\mathrm{d}x}
    = \frac{r}{1 + r^2 x^2}
      + \frac{2-r}{(2-r)^2 + x^2}
      - \frac{2}{1 + x^2}
    = 0.
\end{equation}
Solving this equation for $x^2$ yields a closed-form expression for the frequency at which the phase is extremal. Substituting $x_* = \omega_{*,U}/\kappa$ and rearranging in terms of $r = \kappa \tau_d$, we obtain
\begin{equation}
    \omega_{*, U}^2
    = \kappa^2 \,
      \frac{-r(r-2)(r-1)
      + \sqrt{r(r-2)\bigl\{ (r-1)^4 - 4\bigr\}}}{r(r+1)}.
\end{equation}
A real solution for $\omega_{*,U}$ exists only when the expression under the square root is positive and the right-hand side is non-negative. 
Simple, but somewhat tedious calculation reveals the extremum $\omega_{*,U}$ exists for $1 < r < 2$ and $r > 3$.

The limiting behavior of $H_C^U(\omega)$ is similar to that of $H_C^{w_0}(\omega)$. 
In the quasi-static limit $\omega \to 0$, we find
\begin{align}
    \lim_{\omega \to 0} H_C^U(\omega)
    &= \frac{1}{(\tau_d \kappa)^2} \,
       \frac{\tau_d^{-1}}{\kappa^2} \,
       \bigl(-\tau_d \kappa^2 + 2\kappa\bigr) \\
    &= \frac{2 - \tau_d \kappa}{(\tau_d \kappa)^3}.
\end{align}
Recalling that at steady state the sensitivity for $U$ is
\begin{equation}
    f_*^{U}(\nu)
    = \frac{1}{\bigl(1 + \tau_d \nu U\bigr)^2},
\end{equation}
the corresponding effective presynaptic term reads
\begin{equation}
    C_*^{U}(\nu)
    = \nu f_*^{U}(\nu)
    = \frac{\nu}{\bigl(1 + \tau_d \nu U\bigr)^2}.
\end{equation}
A straightforward calculation then shows
\begin{equation}
    \left.\frac{\mathrm{d} C_*^{U}}{\mathrm{d}\nu}\right|_{\nu=\nu_0}
    = \frac{1 - \tau_d \nu_0 U}{\bigl(1 + \tau_d \nu_0 U\bigr)^3}
    = \frac{2 - \tau_d \kappa}{(\tau_d \kappa)^3}
    = \lim_{\omega \to 0} H_C^U(\omega),
\end{equation}
confirming that quasi-static approximation is valid again.

In the opposite limit of fast modulation, $\omega \to \infty$, we obtain
\begin{equation}
    \lim_{\omega \to \infty} H_C^U(\omega)
    = \frac{1}{(\tau_d \kappa)^2}
    = \frac{1}{\bigl(1 + \tau_d \nu_0 U\bigr)^2}
    = f_*^{U},
\end{equation}
again exhibiting that too rapid fluctuations are effectively averaged out and the gain reduces to the constant factor $f_*^{U}$, i.e., the steady-state sensitivity of the synapse to changes in $U$.

\subsubsection{Dependence on oscillation frequency and STP operating point.}
The small-signal responses derived above depend on the biophysical parameters
only through the dimensionless operating point
\begin{equation}
    r := \tau_d \kappa = 1 + \tau_d \nu_0 U
\end{equation}
and the dimensionless frequency
\begin{equation}
    x := \frac{\omega}{\kappa}.
\end{equation}
For weak sinusoidal input $\nu(t) = \nu_0 + \hat{\nu}\cos(\omega t)$
with $\hat{\nu} \ll \nu_0$, the gain and phase of both the sensitivity
$f^Z(t)$ and the effective presynaptic term $C^Z(t) = f^Z(t)\nu(t)$
are completely determined by $(r,x)$; different combinations of
$(\nu_0, U, \tau_d)$ that yield the same $r$ produce identical
frequency responses.

For the baseline-weight component ($Z = w_0$), the qualitative behavior
is independent of $r$.
As shown analytically above, the phase
\begin{equation}
    \phi_{w_0}(\omega)
    = \arg H_C^{w_0}(\omega)
    = \arctan(\omega \tau_d)
      - \arctan\!\left(\frac{\omega}{\kappa}\right)
\end{equation}
vanishes in both the quasi-static ($\omega \to 0$) and
fast-modulation ($\omega \to \infty$) limits, and exhibits a single
positive maximum at
$\omega_{*,w_0} = \sqrt{\tau_d^{-1}\kappa}$.
Thus, $C^{w_0}(t)$ always shows a band-limited phase lead relative to
$\nu(t)$, with the strongest lead at intermediate frequencies where
the zero at $\omega \sim \tau_d^{-1}$ and the pole at
$\omega \sim \kappa$ interact most strongly.
The gain $|H_C^{w_0}(\omega)|$ increases monotonically from the
quasi-static value $(\kappa\tau_d)^{-2}$ to the high-frequency limit
$f_*^{w_0} = 1/(\tau_d \kappa)$, so slow modulations are attenuated,
whereas faster modulations are relatively enhanced.

\begin{figure}[!htbp]
    \centering
    \includegraphics[width=\linewidth]{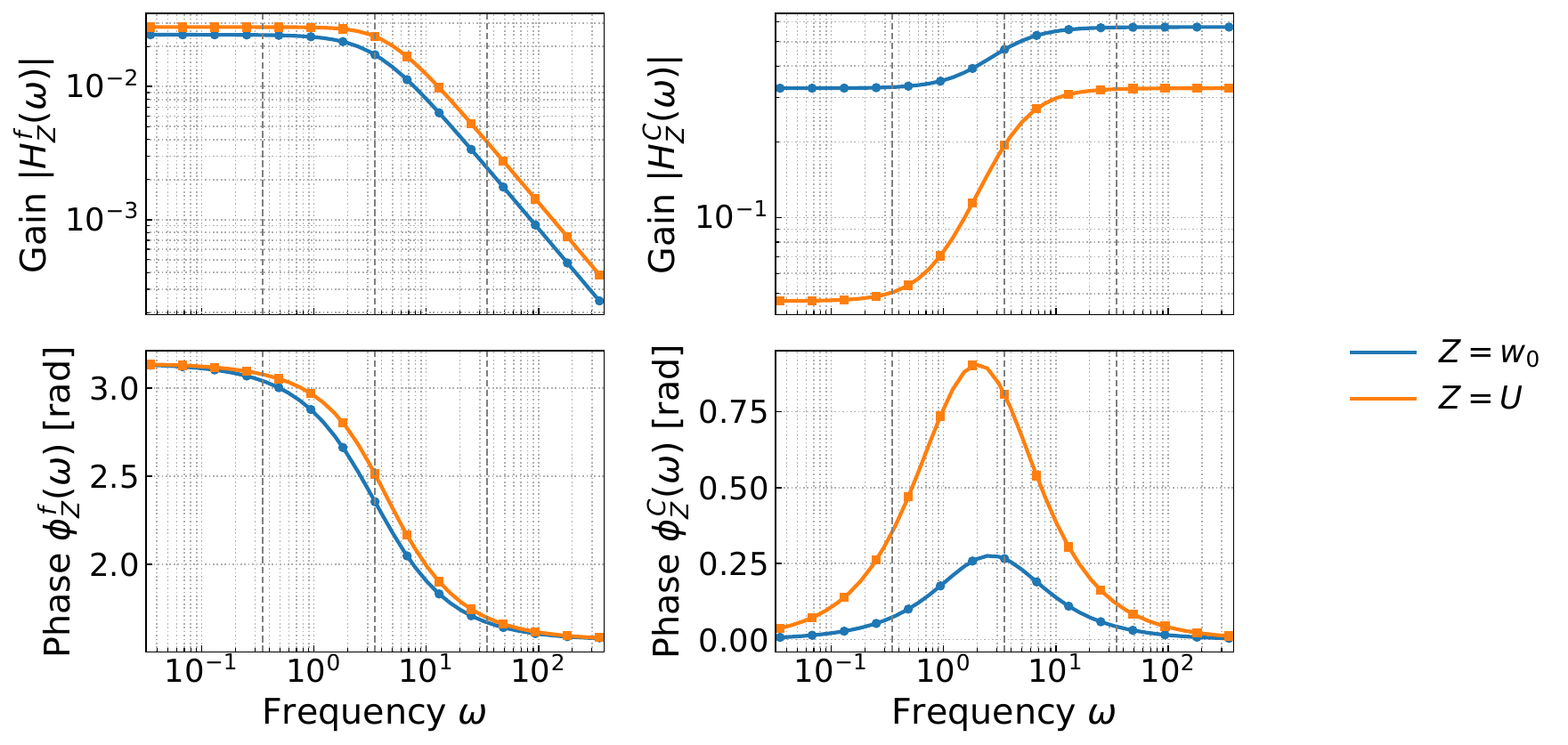}
    \caption[Frequency response for $1 < r < 2$.]%
    {\textbf{Frequency response for $1 < r < 2$.}
    Frequency-domain response of the sensitivity functions $f^{w_0}(t)$
    and $f^{U}(t)$ (left column) and the effective presynaptic terms
    $C^{w_0}(t)$ and $C^{U}(t)$ (right column) 
    with $1 < r = \tau_d \kappa < 2$ (here $r = 1.75$,
    corresponding to $\nu_0 = 10$~Hz, $\tau_d = 0.5$~s, and $U = 0.15$).
    The presynaptic rate is weakly modulated as
    $\nu(t) = \nu_0 + \hat{\nu}\cos(\omega t)$ with
    $\hat{\nu} \ll \nu_0$.
    Top row: gain $|H_f^Z(\omega)|$ and $|H_C^Z(\omega)|$.
    Bottom row: phase $\phi_f^Z(\omega)$ and $\phi_C^Z(\omega)$
    (arguments of the corresponding transfer functions), where positive
    values indicate a phase lead of the output relative to $\nu(t)$.
    Blue curves: $Z = w_0$; orange curves: $Z = U$.
    Solid lines show the analytical linear-response predictions, and
    dots show results from numerical simulations of the full nonlinear
    dynamics.
    In this regime ($1 < r < 2$), both $C^{w_0}$ and $C^{U}$ exhibit a
    single positive phase maximum at intermediate frequencies and
    monotonic increase of the gain with $\omega$.}
    \label{fig:STP-phase-r<2}
\end{figure}

\begin{figure}[!htbp]
    \centering
    \includegraphics[width=\linewidth]{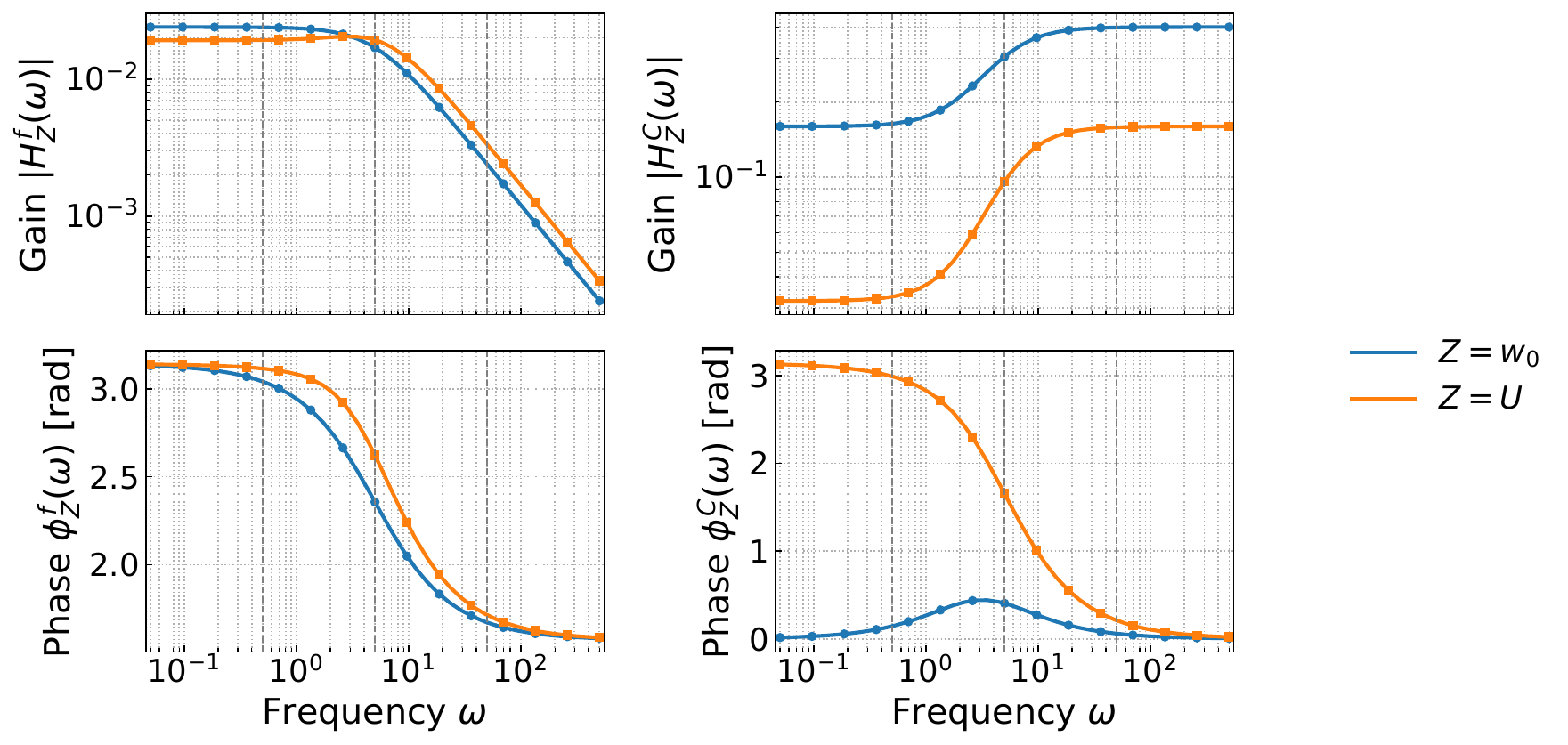}
    \caption[Frequency response for $2 < r < 3$.]%
    {\textbf{Frequency response for $2 < r < 3$.}
    Same layout and conventions as in Fig.~\ref{fig:STP-phase-r<2},
    but with $2 < r = \tau_d \kappa < 3$
    (here $r = 2.5$).
    For $Z = w_0$, the gain and phase closely resemble those in the
    $1 < r < 2$ regime.
    For $Z = U$, however, the phase of both $f^{U}$ and $C^{U}$ is
    close to $\pi$ (almost anti-phase) at very low frequencies and
    decreases monotonically toward $0$ as $\omega$ increases, while the
    gain continues to increase monotonically.
    Thus, in this regime the $U$-dependent contribution to the effective
    presynaptic term inverts very slow input modulations but becomes
    in-phase with the input at high frequencies.}
    \label{fig:STP-phase-2<r<3}
\end{figure}

\begin{figure}[!htbp]
    \centering
    \includegraphics[width=\linewidth]{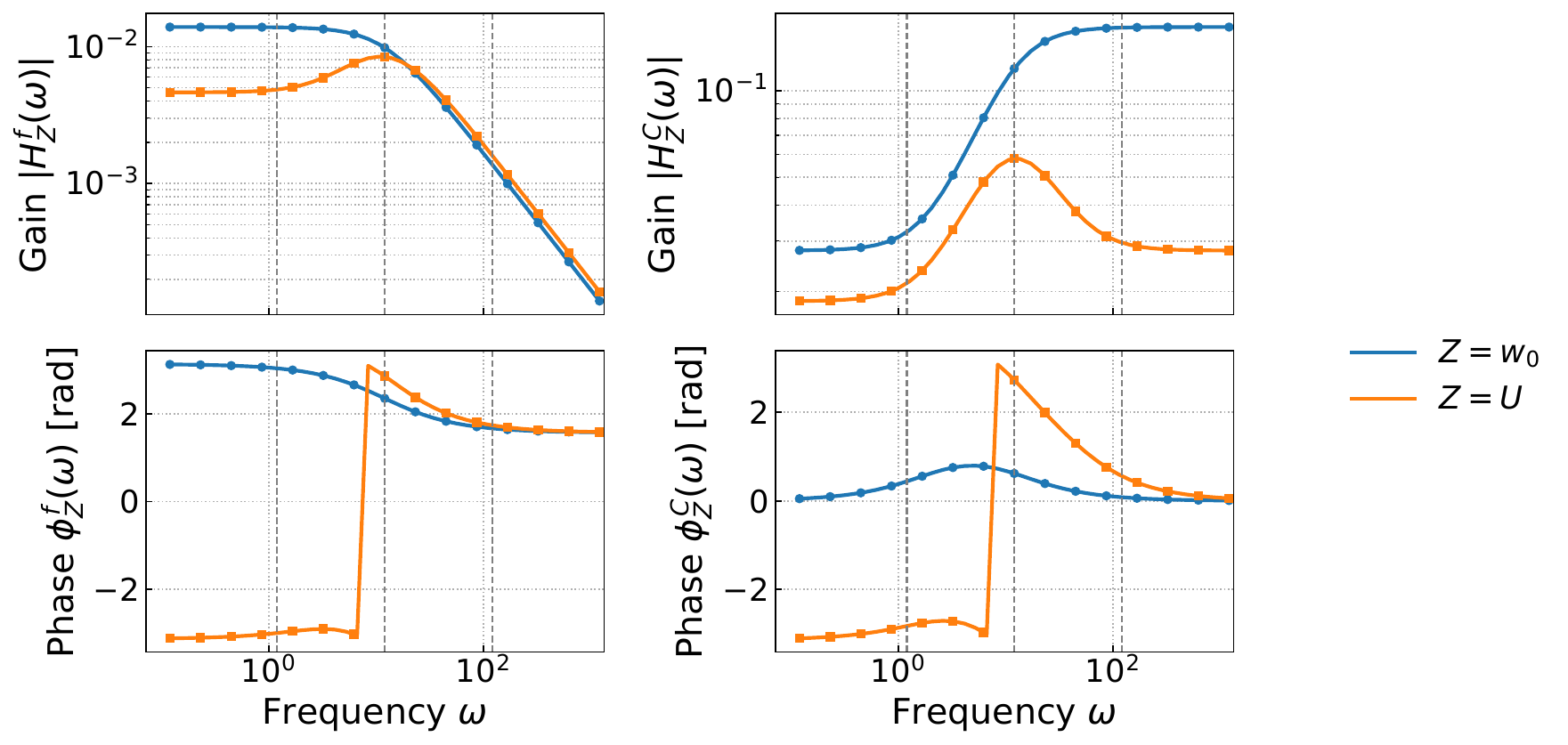}
    \caption[Frequency response for $r > 3$.]%
    {\textbf{Frequency response for $r > 3$.}
    Same layout and conventions as in Fig.~\ref{fig:STP-phase-r<2},
    but for an operating point with $r = \tau_d \kappa > 3$
    (Here $r = 6$ and other parameters are chosen accordingly).
    For $Z = w_0$, the gain and phase again follow the generic pattern
    of a single positive phase maximum and monotonic increase of the
    gain.
    For $Z = U$, the phase curves of $f^{U}$ and $C^{U}$ cross the
    branch cut at $\pm\pi$ in the principal-value representation,
    resulting in an apparent discontinuity.
    When the phase is unwrapped to follow a continuous branch, it
    interpolates from $\phi \approx -\pi$ at very low frequencies,
    through a pronounced positive phase lead at intermediate
    frequencies (peaked at $\omega_{*,U}$), to $\phi \approx 0$ at
    high frequencies.
    In the same regime, the gain $|H_C^{U}(\omega)|$ exhibits a clear
    maximum at intermediate frequencies, indicating that the
    $U$-dependent contribution of the effective presynaptic term is
    band-pass-like and selectively emphasizes the rising phase of
    the input.}
    \label{fig:STP-phase-r>3}
\end{figure}

By contrast, for the release-probability component ($Z = U$) the shape
of the phase response depends qualitatively on $r$.
The analytic expression
\begin{equation}
    \phi_U(\omega)
    = \phi_U(x)
    = \arctan(r x)
      + \arctan\!\left(\frac{x}{2-r}\right)
      - 2\arctan(x)
\end{equation}
yields three distinct regimes (Figs.~\ref{fig:STP-phase-r<2}–\ref{fig:STP-phase-r>3}).
When $1 < r < 2$, the phase behaves similarly to the $w_0$ case:
$\phi_U(\omega)$ is zero at $\omega \to 0$ and $\omega \to \infty$
and exhibits a single positive maximum at an intermediate frequency
$\omega_{*,U}$.
The gain $|H_C^U(\omega)|$ increases monotonically with $\omega$, as in
the $w_0$ case.
For $2 < r < 3$, the phase is approximately $\phi_U(\omega) \approx \pi$
at very low frequencies and decreases monotonically toward $0$ as
$\omega$ increases; thus $C^U(t)$ is nearly in anti-phase with the
slow input and becomes in-phase for rapid modulations.
The gain remains monotonically increasing.
For $r > 3$, the phase curve crosses the branch cut at $\pm\pi$;
when the phase is plotted as a continuous (unwrapped) branch,
$\phi_U(\omega)$ again exhibits an extremum at $\omega_{*,U}$,
but now interpolates from $\phi_U \approx -\pi$ at very low
frequencies to $\phi_U \approx 0$ at high frequencies.
In physiologically relevant ranges of $r$, this means that $C^U (t)$ can be almost perfectly anti-phase with
slow inputs, show a pronounced positive phase lead at intermediate
frequencies, and become nearly synchronous with $\nu(t)$ at high
frequencies.
In the same regime, the gain $|H_C^U(\omega)|$ exhibits a clear
maximum at intermediate frequencies, indicating a band-pass-like
sensitivity of $C^U(t)$ to the onset of rate changes.

Comparing the two components, the effective presynaptic term driven by
$w_0$ always has the larger gain, $|H_C^{w_0}(\omega)| >
|H_C^{U}(\omega)|$, whereas the phase lead (after unwrapping the branch
cut at $\pm\pi$) is systematically larger for $U$ than for $w_0$ over
the relevant frequency range.
Finally, increasing $r$ reduces the gain at all frequencies for both
parameters, reflecting the stronger overall depression at higher firing
rates and release probabilities, while at the same time it enhances
the magnitude of the phase shift: depression is engaged earlier within
each cycle, so the peak of $C^Z(t)$ shifts toward the rising flank of
$\nu(t)$.

Finally, to verify the robustness of these conclusions, we also examined the
case of finite-amplitude modulation, $\hat{\nu} = \nu_0$.
Although the numerical results then deviate quantitatively from the
linear-response predictions, the overall frequency dependence of both
gain and phase is essentially unchanged: the relative ordering of
$|H_C^{w_0}|$ and $|H_C^{U}|$, the presence or absence of an
intermediate-frequency maximum, and the characteristic phase lead/lag
patterns across $r$ remain intact (Fig.~\ref{fig:STP-phase-large-perturb}).

\begin{figure}[!htbp]
    \centering
    \includegraphics[width=\linewidth]{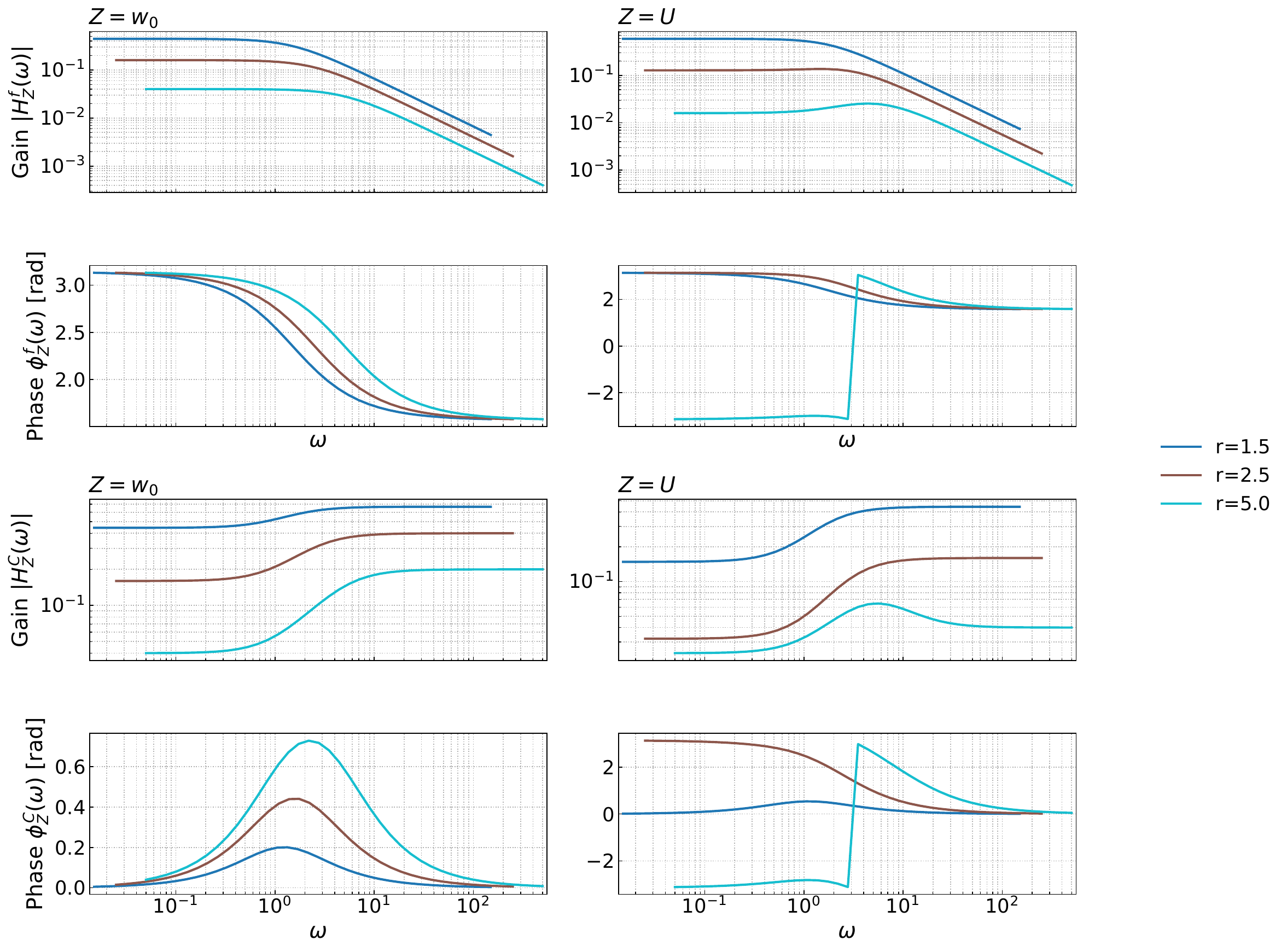}
    \caption[Comparison across different operating points $r$.]%
    {\textbf{Comparison across different operating points $r$.}
    Frequency responses of the sensitivity functions and
    effective presynaptic terms for several values of
    the operating point $r = \tau_d \kappa$ (different curves,
    indicated in the legend).
    Upper panels: gain $|H_f^Z(\omega)|$ and $|H_C^Z(\omega)|$ for
    $Z \in \{w_0,U\}$. 
    Lower panels: corresponding phases
    $\phi_f^Z(\omega)$ and $\phi_C^Z(\omega)$.
    Left column: $Z = w_0$. Right column: $Z=U$.
    For both parameters, increasing $r$ decreases the gain at all
    frequencies, reflecting stronger overall short-term depression at
    higher firing rates and release probabilities.
    At the same time, the magnitude of the phase shift (relative lead
    or lag) increases over a broad range of intermediate frequencies,
    indicating that depression is recruited earlier within each cycle
    when the synapse operates in a more strongly depressed regime.}
    \label{fig:STP-phase-multi-r}
\end{figure}

\begin{figure}[!htbp]
    \centering
    \includegraphics[width=\linewidth]{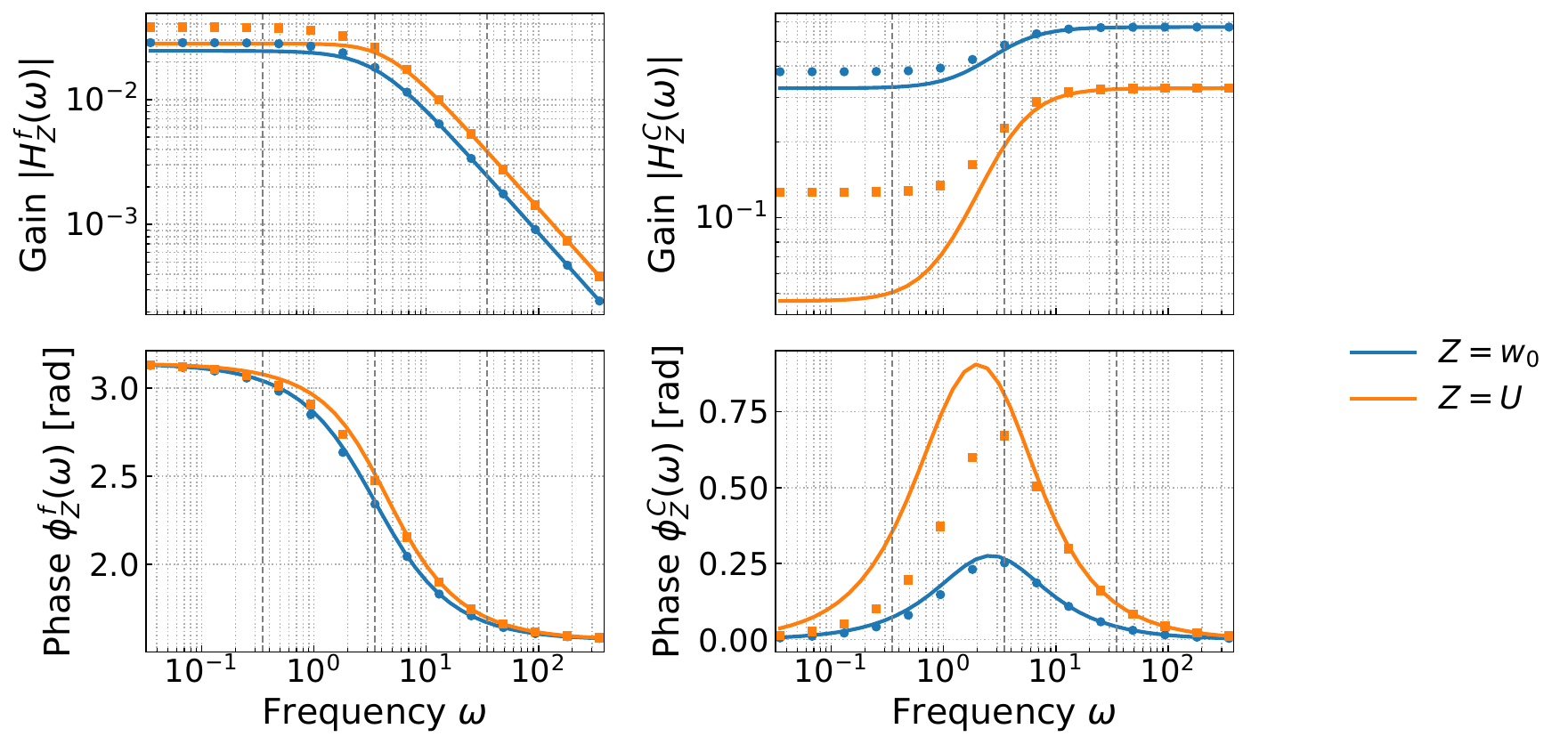}
    \caption[Effect of finite-amplitude modulation.]%
    {\textbf{Effect of finite-amplitude modulation.}
    Frequency responses of the sensitivity and effective presynaptic
    terms for a larger modulation amplitude,
    $\nu(t) = \nu_0 + \hat{\nu}\cos(\omega t)$ with
    $\hat{\nu} = \nu_0 = 10$~Hz.
    The parameters match those used in Fig.~\ref{fig:STP-phase-response-C} and in
    Figs.~\ref{fig:STP-phase-r<2}–\ref{fig:STP-phase-multi-r}.
    Panel layout and color conventions are identical to the previous
    figures: blue for $Z = w_0$, orange for $Z = U$, solid lines for
    analytical predictions based on linear response, and dots for
    numerical simulations.
    Although the larger modulation amplitude leads to quantitative
    deviations from the linear-response curves, the qualitative
    behavior of both gain and phase is unchanged:
    the relative ordering of $|H_C^{w_0}|$ and $|H_C^{U}|$, the
    presence or absence of an intermediate-frequency maximum in the
    gain, and the characteristic phase-lead patterns across different
    operating points $r$ remain essentially the same.}
    \label{fig:STP-phase-large-perturb}
\end{figure}

\clearpage
\section{Additional Figures}

\begin{figure}[!htbp]
    \centering
    \includegraphics[width=\linewidth]{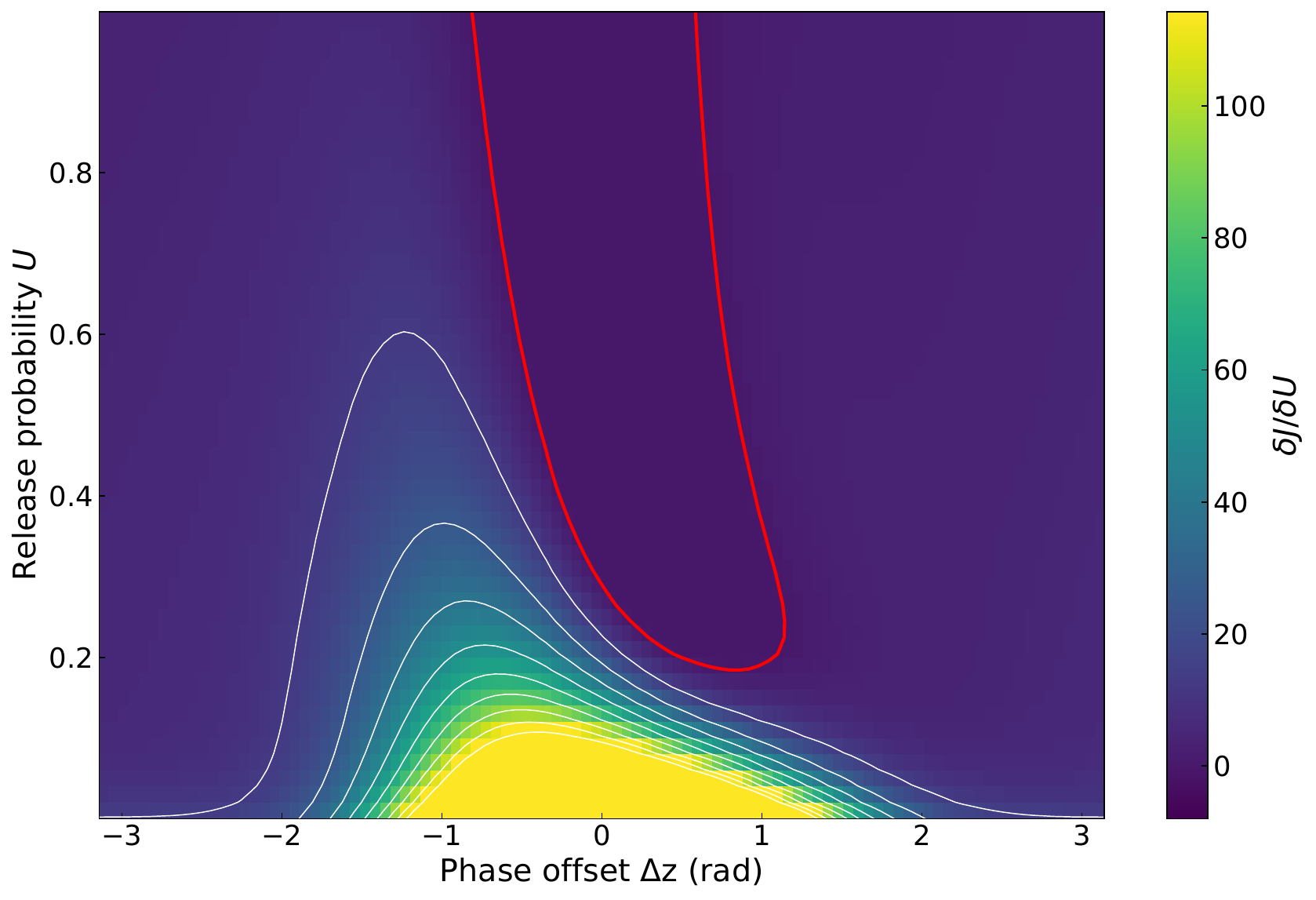}
    \caption[Gradient of Fisher information with respect to $U(\Delta z)$.]
    {\textbf{Gradient of Fisher information with respect to $U(\Delta z)$.} 
    Heat map of $\frac{\delta J}{\delta U(\Delta z)}$ as a function of phase difference $\Delta z$ and release probability $U$.
    White contour lines indicate optimal solutions under the constant-sum constraint.
    The red contour marks the boundary where $\frac{\delta J}{\delta U(\Delta z)} = 0$.
    Parameters are identical to Figure~\ref{fig:pre-post_factors}.}
    \label{fig:grad-U}
\end{figure}

\begin{figure}[!htbp]
\centering
    \includegraphics[width=\linewidth]{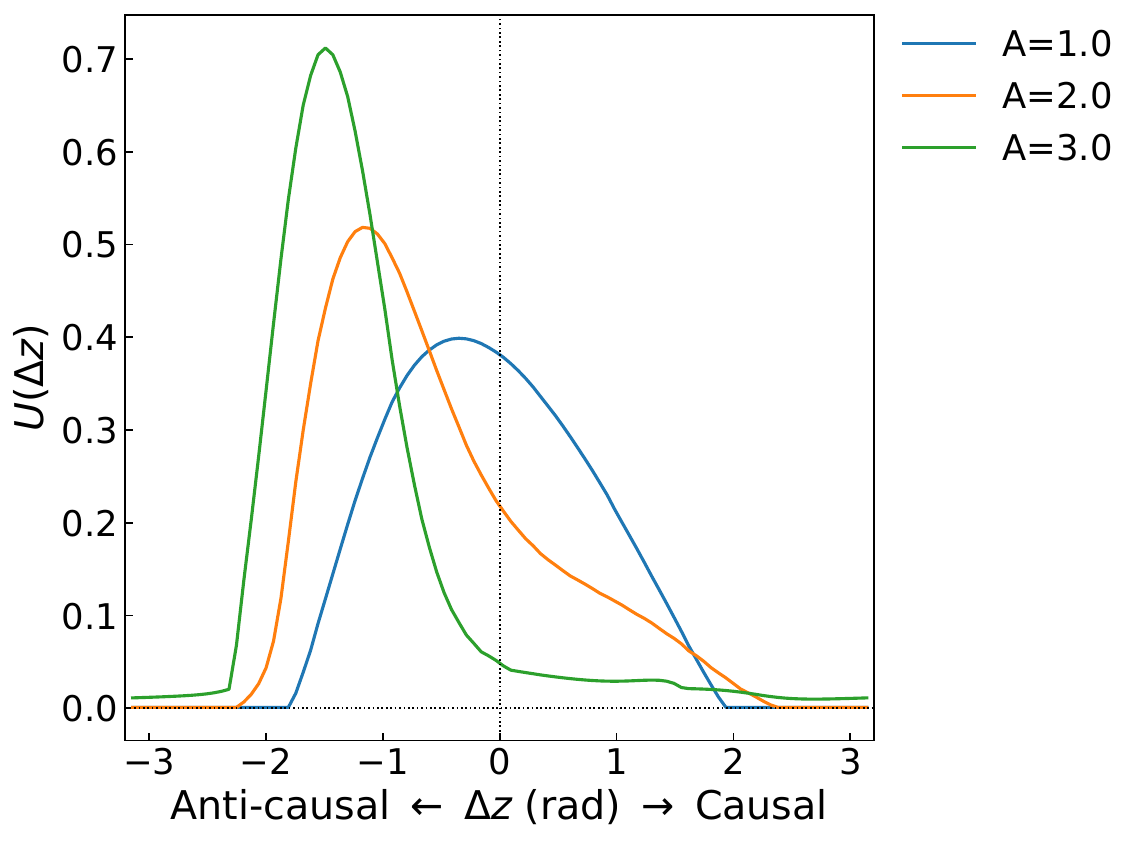}
    \caption[Optimal $U(\Delta z)$ under different input amplitudes.]
    {\textbf{Optimal $U(\Delta z)$ under different input amplitudes.}
    Optimization results for $U(\Delta z)$ with different amplitudes of input $h(z,t) = A[\cos \omega t - \cos \theta_c]$.
    Stronger input results in more skewed connectivity.
    Blue: $A = 1.0$. Orange: $A = 2.0$. Green: $A = 3.0$. Other parameters are identical to the figures in the subsection \ref{subsec:STP-traveling-wave}.
    }
    \label{fig:optimize-U-supple-amplitudes}
\end{figure}

\begin{figure*}[!htbp]
    \centering
    \includegraphics[width=0.8\linewidth]{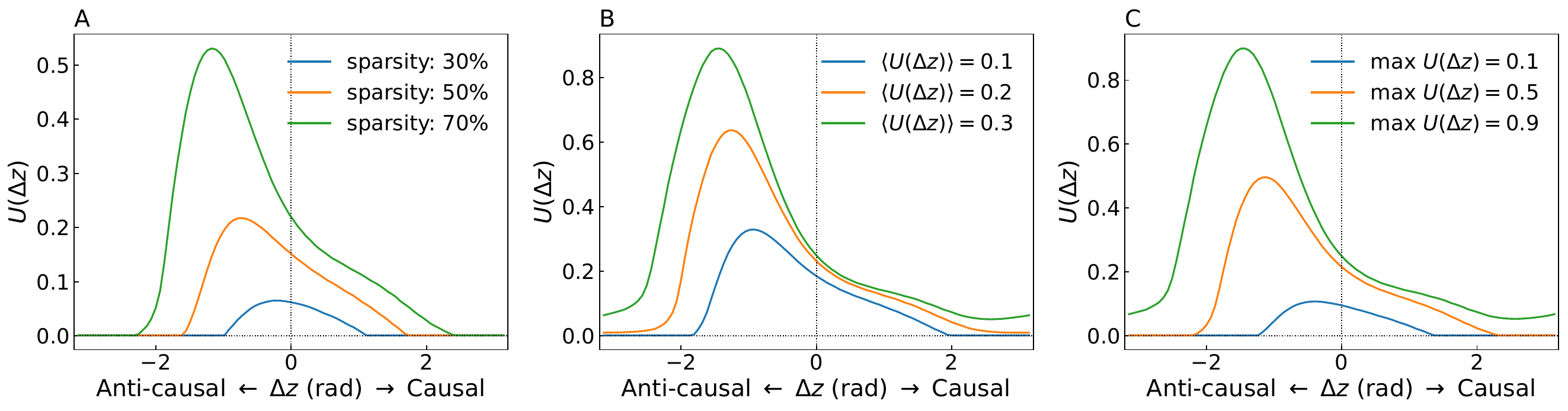}
    \caption[Optimal $U(\Delta z)$ under different constraints.]
    {\textbf{Optimal $U(\Delta z)$ under different constraints.}
    Optimization results for $U(\Delta z)$ under various restrictions.
    The overall trend remains unchanged by the type of restrictions.
    \textbf{A}. different sparsity level. 
    \textbf{B}. different mean $U(\Delta z)$. 
    \textbf{C}. different $\max U(\Delta z)$.
    Other parameters are identical to the figures in the subsection \ref{subsec:STP-traveling-wave}.
    }
    \label{fig:optimize-U-supple-constraints}
\end{figure*}

\begin{figure*}[!htbp]
    \centering
    \includegraphics[width=0.8\linewidth]{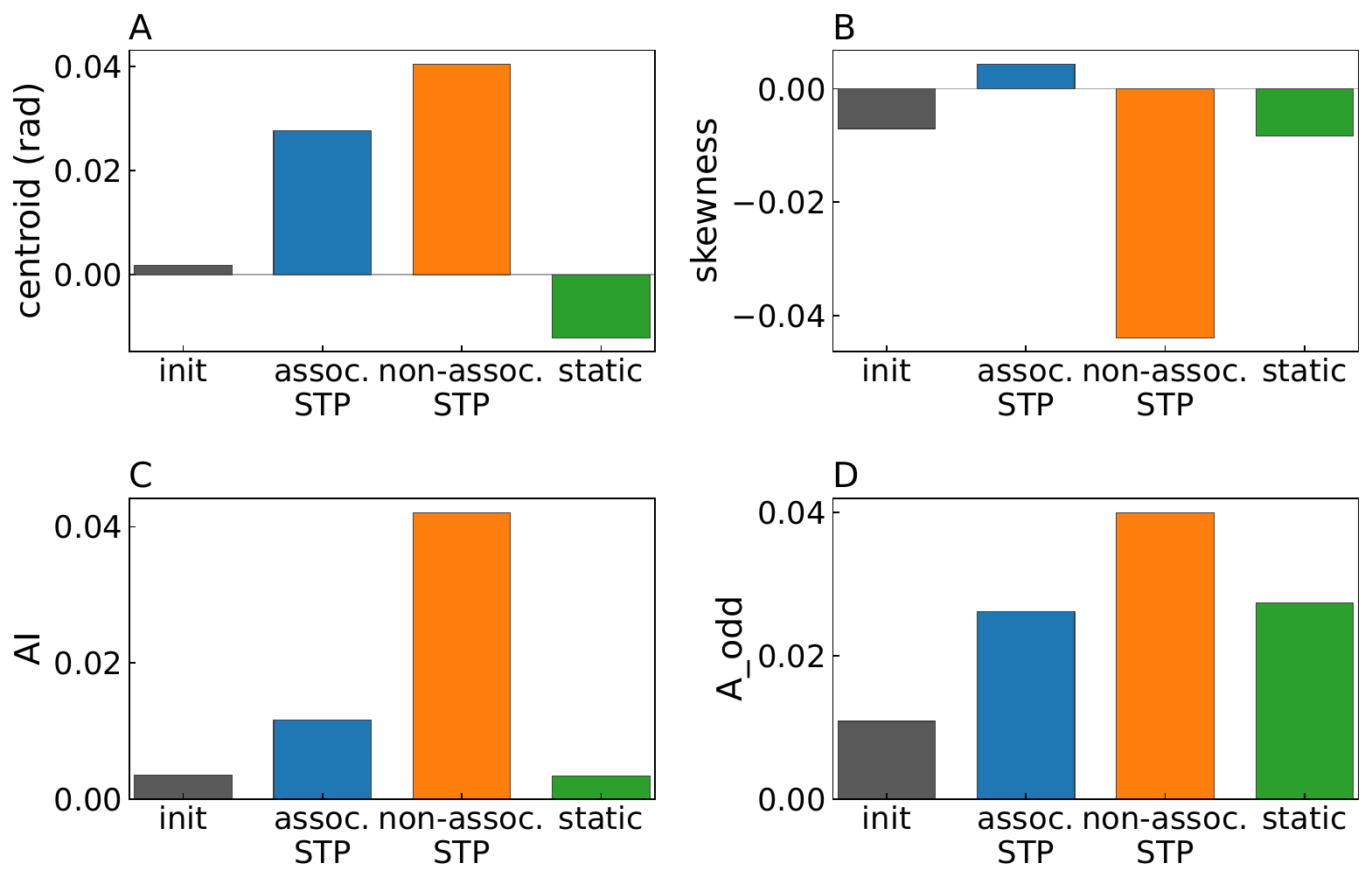}
    \caption[Quantifying temporal asymmetry of stimulus-evoked responses.]
    {\textbf{Quantifying temporal asymmetry of stimulus-evoked responses.}
    Temporal asymmetry metrics computed from the phase-aligned rate $r(\phi)$ in Figure~\ref{fig:evoked-activity}A for the four conditions (initial, associative STP, non-associative STP, and static synapses; same color code).
    \textbf{A}. Centroid phase $\mu = \sum_i r_i\phi_i / \sum_i r_i$.
    \textbf{B}. Skewness $m_3/m_2^{3/2}$, with $m_k = \sum_i r_i(\phi_i-\mu)^k / \sum_i r_i$.
    \textbf{C}. Area index $\mathrm{AI}=(M_+ - M_-)/R$, where $M_+ = \sum_{\phi_i>0} r_i$, $M_- = \sum_{\phi_i<0} r_i$, and $R=\sum_i r_i$.
    \textbf{D}. Odd ratio $A_{\mathrm{odd}}=\lVert r_{\mathrm{odd}}\rVert_2/\lVert r\rVert_2$, where $r_{\mathrm{odd}}(\phi)=(r(\phi)-r(-\phi))/2$.
    Metrics in \textbf{A-C} report direction (later vs.\ earlier phases within the stimulus epoch), whereas \textbf{D} quantifies asymmetry magnitude independent of direction.}
    \label{fig:asymmetry-metrics}
\end{figure*}

\begin{figure*}[!htbp]
    \centering
    \includegraphics[width=0.8\linewidth]{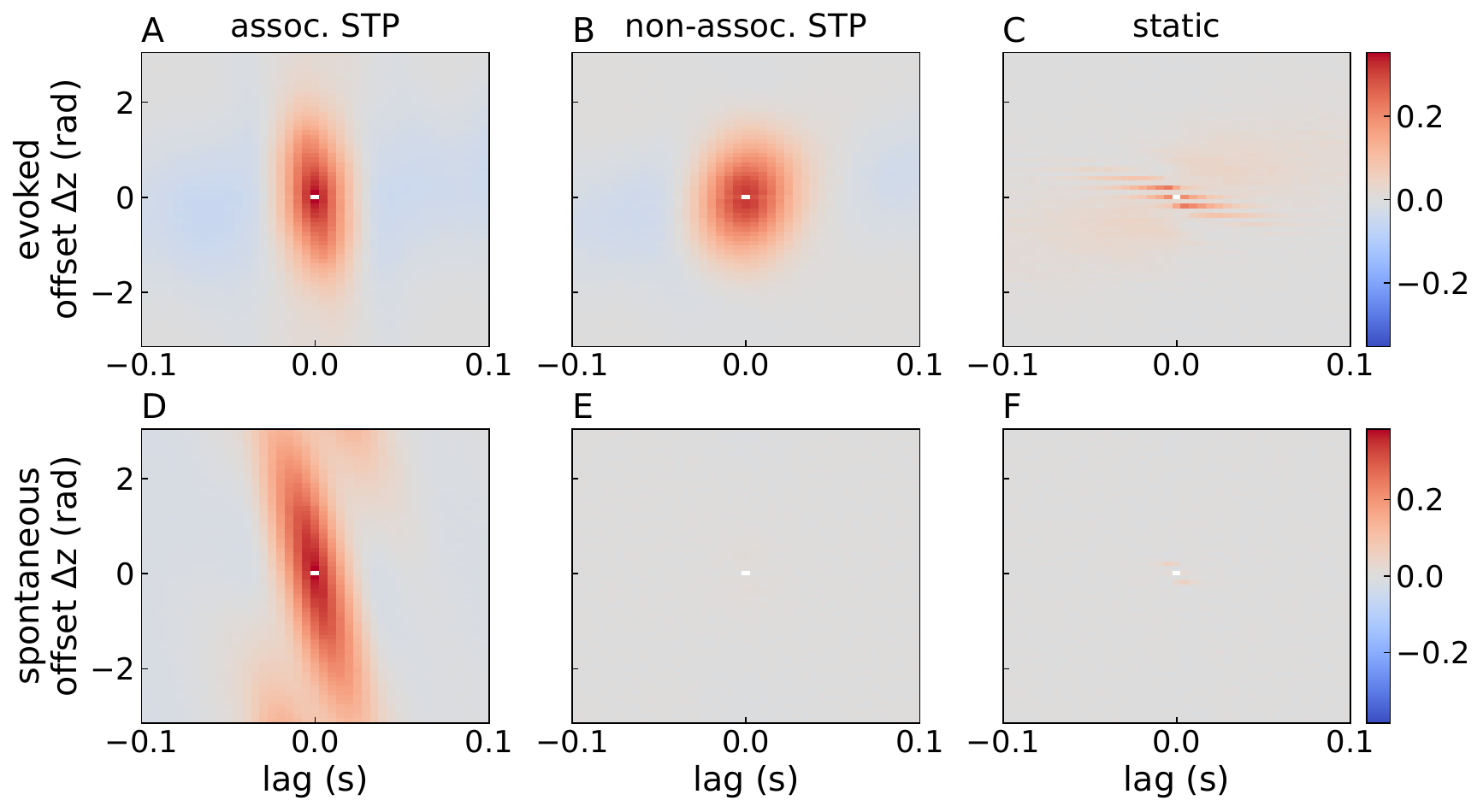}
    \caption[Cross-correlograms across STP conditions]{
    \textbf{Cross-correlograms across STP conditions during evoked and spontaneous activity.}
    Trial-shuffle-corrected pairwise spike-train cross-correlograms after Fisher-information optimization,
    shown as a function of ring offset $\Delta z$ and time lag $\tau$.
    \textbf{A--C.} During stimulus presentation.
    \textbf{D--F.} During spontaneous activity.
    \textbf{A,D.} Associative STP.
    \textbf{B,E.} Non-associative STP.
    \textbf{C,F.} Static synapses.
    During spontaneous activity (\textbf{D--F}), the traveling-wave drive is removed and the network receives a
    uniform background input $h_{\mathrm{bg}}=0.5$.
    The zero-lag autocorrelation bin at $\Delta z=0$ is omitted.
    Simulation, learning, and resource-constraint parameters are identical to
    Figure~\ref{fig:evoked-activity}.}
   \label{fig:cross-correlograms}
\end{figure*}

\clearpage
\bibliography{Thesis}

\end{document}